\RequirePackage{lineno}
\documentclass[aps,prl,twocolumn,showpacs,superscriptaddress,groupedaddress]{revtex4}  
\usepackage{graphicx}  
\usepackage{dcolumn}   
\usepackage{bm}        
\usepackage{amssymb}   
\usepackage{multirow}
\usepackage{epsfig}

\newcommand{\zgamma}{\ensuremath{Z/\gamma^{*}}}
\newcommand{\stwcenter}{0.2304}

\newcommand{\stwstaterr}{0.0008}
\newcommand{\stwsysterr}{0.0006}
\newcommand{\stwsysterrpdf}{0.0005}
\newcommand{\stwsysterrscale}{0.0003}

\newcommand{\thetalep}{\ensuremath{\theta_{\text{eff}}^{\text{$\ell$}}}}
\newcommand{\stwefflep}{\ensuremath{\sin^2\theta_{\text{eff}}^{\text{$\ell$}}}}

\newcommand{\stweffu}{\ensuremath{\sin^2\theta_{\text{eff}}^{\it{u}}}}
\newcommand{\stweffd}{\ensuremath{\sin^2\theta_{\text{eff}}^{\it{d}}}}
\newcommand{\stwefffermion}{\ensuremath{\sin^2\theta_{\text{eff}}^{\it{f}}}}
\newcommand{\metpaul}{\mbox{$\rlap{\kern0.15em/}E_T$}}
\begin{document}

\hspace{5.2in} \mbox{FERMILAB-PUB-11-190-E}

\title{Measurement of $\boldsymbol{\stwefflep}$ and $\boldsymbol{Z}$-light quark couplings using the forward-backward
charge asymmetry in $\boldsymbol{p\bar{p} \rightarrow Z/\gamma^{*} \rightarrow e^{+}e^{-}}$ events with ${\cal \boldsymbol{L}}\boldsymbol{=5.0}$ fb$\boldsymbol{^{-1}}$ at $\boldsymbol{\sqrt{s}=1.96}$~TeV}

%
\affiliation{Universidad de Buenos Aires, Buenos Aires, Argentina}
\affiliation{LAFEX, Centro Brasileiro de Pesquisas F{\'\i}sicas, Rio de Janeiro, Brazil}
\affiliation{Universidade do Estado do Rio de Janeiro, Rio de Janeiro, Brazil}
\affiliation{Universidade Federal do ABC, Santo Andr\'e, Brazil}
\affiliation{Instituto de F\'{\i}sica Te\'orica, Universidade Estadual Paulista, S\~ao Paulo, Brazil}
\affiliation{Simon Fraser University, Vancouver, British Columbia, and York University, Toronto, Ontario, Canada}
\affiliation{University of Science and Technology of China, Hefei, People's Republic of China}
\affiliation{Universidad de los Andes, Bogot\'{a}, Colombia}
\affiliation{Charles University, Faculty of Mathematics and Physics, Center for Particle Physics, Prague, Czech Republic}
\affiliation{Czech Technical University in Prague, Prague, Czech Republic}
\affiliation{Center for Particle Physics, Institute of Physics, Academy of Sciences of the Czech Republic, Prague, Czech Republic}
\affiliation{Universidad San Francisco de Quito, Quito, Ecuador}
\affiliation{LPC, Universit\'e Blaise Pascal, CNRS/IN2P3, Clermont, France}
\affiliation{LPSC, Universit\'e Joseph Fourier Grenoble 1, CNRS/IN2P3, Institut National Polytechnique de Grenoble, Grenoble, France}
\affiliation{CPPM, Aix-Marseille Universit\'e, CNRS/IN2P3, Marseille, France}
\affiliation{LAL, Universit\'e Paris-Sud, CNRS/IN2P3, Orsay, France}
\affiliation{LPNHE, Universit\'es Paris VI and VII, CNRS/IN2P3, Paris, France}
\affiliation{CEA, Irfu, SPP, Saclay, France}
\affiliation{IPHC, Universit\'e de Strasbourg, CNRS/IN2P3, Strasbourg, France}
\affiliation{IPNL, Universit\'e Lyon 1, CNRS/IN2P3, Villeurbanne, France and Universit\'e de Lyon, Lyon, France}
\affiliation{III. Physikalisches Institut A, RWTH Aachen University, Aachen, Germany}
\affiliation{Physikalisches Institut, Universit{\"a}t Freiburg, Freiburg, Germany}
\affiliation{II. Physikalisches Institut, Georg-August-Universit{\"a}t G\"ottingen, G\"ottingen, Germany}
\affiliation{Institut f{\"u}r Physik, Universit{\"a}t Mainz, Mainz, Germany}
\affiliation{Ludwig-Maximilians-Universit{\"a}t M{\"u}nchen, M{\"u}nchen, Germany}
\affiliation{Fachbereich Physik, Bergische Universit{\"a}t Wuppertal, Wuppertal, Germany}
\affiliation{Panjab University, Chandigarh, India}
\affiliation{Delhi University, Delhi, India}
\affiliation{Tata Institute of Fundamental Research, Mumbai, India}
\affiliation{University College Dublin, Dublin, Ireland}
\affiliation{Korea Detector Laboratory, Korea University, Seoul, Korea}
\affiliation{CINVESTAV, Mexico City, Mexico}
\affiliation{FOM-Institute NIKHEF and University of Amsterdam/NIKHEF, Amsterdam, The Netherlands}
\affiliation{Radboud University Nijmegen/NIKHEF, Nijmegen, The Netherlands}
\affiliation{Joint Institute for Nuclear Research, Dubna, Russia}
\affiliation{Institute for Theoretical and Experimental Physics, Moscow, Russia}
\affiliation{Moscow State University, Moscow, Russia}
\affiliation{Institute for High Energy Physics, Protvino, Russia}
\affiliation{Petersburg Nuclear Physics Institute, St. Petersburg, Russia}
\affiliation{Instituci\'{o} Catalana de Recerca i Estudis Avan\c{c}ats (ICREA) and Institut de F\'{i}sica d'Altes Energies (IFAE), Barcelona, Spain}
\affiliation{Stockholm University, Stockholm and Uppsala University, Uppsala, Sweden}
\affiliation{Lancaster University, Lancaster LA1 4YB, United Kingdom}
\affiliation{Imperial College London, London SW7 2AZ, United Kingdom}
\affiliation{The University of Manchester, Manchester M13 9PL, United Kingdom}
\affiliation{University of Arizona, Tucson, Arizona 85721, USA}
\affiliation{University of California Riverside, Riverside, California 92521, USA}
\affiliation{Florida State University, Tallahassee, Florida 32306, USA}
\affiliation{Fermi National Accelerator Laboratory, Batavia, Illinois 60510, USA}
\affiliation{University of Illinois at Chicago, Chicago, Illinois 60607, USA}
\affiliation{Northern Illinois University, DeKalb, Illinois 60115, USA}
\affiliation{Northwestern University, Evanston, Illinois 60208, USA}
\affiliation{Indiana University, Bloomington, Indiana 47405, USA}
\affiliation{Purdue University Calumet, Hammond, Indiana 46323, USA}
\affiliation{University of Notre Dame, Notre Dame, Indiana 46556, USA}
\affiliation{Iowa State University, Ames, Iowa 50011, USA}
\affiliation{University of Kansas, Lawrence, Kansas 66045, USA}
\affiliation{Kansas State University, Manhattan, Kansas 66506, USA}
\affiliation{Louisiana Tech University, Ruston, Louisiana 71272, USA}
\affiliation{Boston University, Boston, Massachusetts 02215, USA}
\affiliation{Northeastern University, Boston, Massachusetts 02115, USA}
\affiliation{University of Michigan, Ann Arbor, Michigan 48109, USA}
\affiliation{Michigan State University, East Lansing, Michigan 48824, USA}
\affiliation{University of Mississippi, University, Mississippi 38677, USA}
\affiliation{University of Nebraska, Lincoln, Nebraska 68588, USA}
\affiliation{Rutgers University, Piscataway, New Jersey 08855, USA}
\affiliation{Princeton University, Princeton, New Jersey 08544, USA}
\affiliation{State University of New York, Buffalo, New York 14260, USA}
\affiliation{Columbia University, New York, New York 10027, USA}
\affiliation{University of Rochester, Rochester, New York 14627, USA}
\affiliation{State University of New York, Stony Brook, New York 11794, USA}
\affiliation{Brookhaven National Laboratory, Upton, New York 11973, USA}
\affiliation{Langston University, Langston, Oklahoma 73050, USA}
\affiliation{University of Oklahoma, Norman, Oklahoma 73019, USA}
\affiliation{Oklahoma State University, Stillwater, Oklahoma 74078, USA}
\affiliation{Brown University, Providence, Rhode Island 02912, USA}
\affiliation{University of Texas, Arlington, Texas 76019, USA}
\affiliation{Southern Methodist University, Dallas, Texas 75275, USA}
\affiliation{Rice University, Houston, Texas 77005, USA}
\affiliation{University of Virginia, Charlottesville, Virginia 22901, USA}
\affiliation{University of Washington, Seattle, Washington 98195, USA}
\author{V.M.~Abazov} \affiliation{Joint Institute for Nuclear Research, Dubna, Russia}
\author{B.~Abbott} \affiliation{University of Oklahoma, Norman, Oklahoma 73019, USA}
\author{B.S.~Acharya} \affiliation{Tata Institute of Fundamental Research, Mumbai, India}
\author{M.~Adams} \affiliation{University of Illinois at Chicago, Chicago, Illinois 60607, USA}
\author{T.~Adams} \affiliation{Florida State University, Tallahassee, Florida 32306, USA}
\author{G.D.~Alexeev} \affiliation{Joint Institute for Nuclear Research, Dubna, Russia}
\author{G.~Alkhazov} \affiliation{Petersburg Nuclear Physics Institute, St. Petersburg, Russia}
\author{A.~Alton$^{a}$} \affiliation{University of Michigan, Ann Arbor, Michigan 48109, USA}
\author{G.~Alverson} \affiliation{Northeastern University, Boston, Massachusetts 02115, USA}
\author{G.A.~Alves} \affiliation{LAFEX, Centro Brasileiro de Pesquisas F{\'\i}sicas, Rio de Janeiro, Brazil}
% ...
%
%
\collaboration{The D0 Collaboration\footnote{with visitors from
$^{a}$Augustana College, Sioux Falls, SD, USA,
$^{b}$The University of Liverpool, Liverpool, UK,
$^{c}$SLAC, Menlo Park, CA, USA,
$^{d}$University College London, London, UK,
$^{e}$Centro de Investigacion en Computacion - IPN, Mexico City, Mexico,
$^{f}$ECFM, Universidad Autonoma de Sinaloa, Culiac\'an, Mexico,
and 
$^{g}$Universit{\"a}t Bern, Bern, Switzerland.
}} \noaffiliation
\vskip 0.25cm

\date{June 28, 2011}
\begin{abstract}
We measure the mass dependence of the forward-backward charge asymmetry in 157,553
$p\bar{p} \rightarrow Z/\gamma^{*} \rightarrow e^+e^-$ interactions, 
corresponding to 5.0 fb$^{-1}$ of integrated luminosity collected by the D0 experiment at the 
Fermilab Tevatron Collider at $\sqrt{s}=1.96$ TeV. The effective weak
mixing angle ($\thetalep$) from this process involving predominantly the first
generation of quarks is extracted as $\stwefflep = 0.2309 \pm
\stwstaterr ~(\mbox{stat.}) \pm \stwsysterr~(\mbox{syst.})$. We also
present the most precise direct measurement of the vector and
axial-vector couplings of $u$ and $d$ quarks to the $Z$ boson.
\end{abstract}

\maketitle

\section{Introduction}
\begin{figure}[htbp]
\epsfig{file=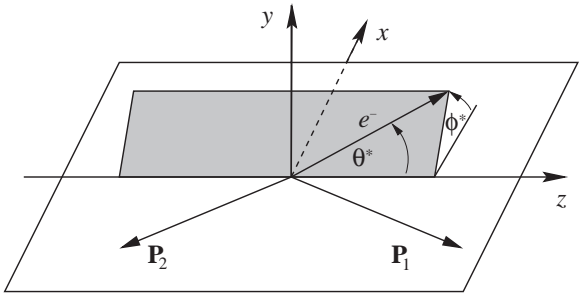, scale = 0.3} \caption{\small
The Collins-Soper reference frame. The bisector of the proton beam
momentum {\cal $\boldsymbol{P_{1}}$} and the negative of the anti-proton beam
momentum {\cal $\boldsymbol{-P_{2}}$} are used to measure the angle $\theta^*$.
The momenta {\cal $\boldsymbol{P_{1}}$} and {\cal $\boldsymbol{P_{2}}$} are measured in the $e^+e^-$ rest
frame.} \label{fig:csframe_new}
\end{figure}

Electron-positron pairs ($e^{+}e^{-}$) can be produced through the
Drell-Yan process over a large invariant mass range at the Fermilab
Tevatron Collider. In the standard model (SM) of particle physics,
the process occurs to first order via $q\bar{q}$ annihilation into a
real (or virtual) $Z$ boson or a virtual photon ($\gamma^*$). While
the coupling of a fermion ($f$) to the photon is purely a vector
coupling, the coupling of the same fermion to the $Z$ boson has both
vector ($g_V^f$) and axial-vector ($g_A^f$) components:
\begin{equation}
 \begin{array}{l}
  \displaystyle g_V^f = I_3^f - 2q_f\cdot\sin^2\theta_W, \\
  \displaystyle g_A^f = I_3^f,
 \end{array}
\label{eqn:ZffVA2}
\end{equation}
where $I_3^f$ and $q_f$ are the third component of the weak isospin and the
charge of the fermion, and $\theta_W$ is the weak mixing angle~\cite{pdg}. The
presence of both vector and axial-vector couplings gives rise to an
asymmetry in the distribution of the polar angle $\theta^*$ of the
negatively charged lepton relative to the incoming quark direction 
in the rest frame of the lepton pair. To minimize the
effect of the unknown transverse momentum of the incoming quarks, we
calculate $\theta^*$ in the Collins-Soper reference
frame~\cite{cs_frame} as
\begin{equation}
 \cos\theta^* = \frac{2}{|Q|\sqrt{Q^2+Q^2_T}}(P^+_{l} P^-_{\bar{l}} - P^-_{l} P^+_{\bar{l}}),
\end{equation}
where {\cal $Q$} ({\cal $Q_T$}) is the four momentum (transverse
momentum) of the lepton pair, and {\cal $P_{l}$} and {\cal
$P_{\bar{l}}$} are the four momenta of the lepton and anti-lepton,
respectively. They are measured in the lab frame, and the momenta $P^{\pm}_{l}$ are defined as 
\begin{equation}
 P^{\pm}_{l} = \frac{1}{\sqrt{2}}(P^0_l \pm P^3_l),
\end{equation}
where {\cal $P^0_l$} and {\cal $P^3_l$} are the energy and the longitudinal component of the lepton
momentum, respectively. In the Collins-Soper frame, the polar axis is defined as the bisector of
the proton beam momentum {\cal $\boldsymbol{P_1}$} and the negative of the anti-proton
beam momentum, {\cal $\boldsymbol{-P_2}$}, when the proton and anti-proton are boosted into
the rest frame of the lepton pair, as shown in
Fig.~\ref{fig:csframe_new}~\cite{cs_frame_plot}.

Events with electron $\cos \theta^*>0$ are classified as forward ($F$), and
those with electron $\cos \theta^*<0$ are classified as backward ($B$). The
forward-backward charge asymmetry, $A_{FB}$, is defined by
\begin{equation}
  A_{FB}=\frac{\sigma_F-\sigma_B}{\sigma_F+\sigma_B},
\end{equation}
where $\sigma_{F}$ and $\sigma_{B}$ are the cross 
sections for forward and backward processes, respectively.

\begin{figure}[htbp]
\epsfig{file=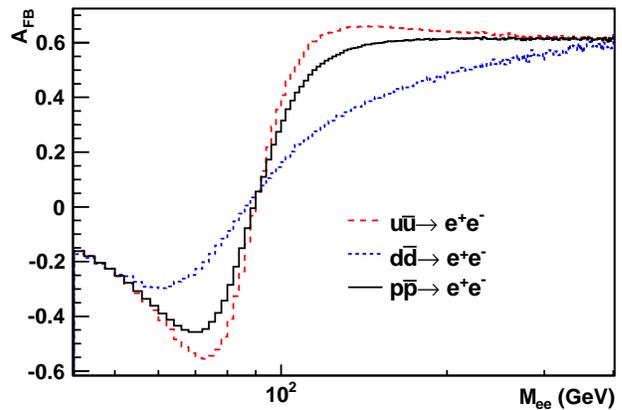, scale = 0.45}
\caption{\small [color online] The SM LO $A_{FB}$ prediction as a function of the dielectron invariant mass for
$u\bar{u} \rightarrow e^+e^-$, $d\bar{d} \rightarrow e^+e^-$, and
$p\bar{p} \rightarrow e^+e^-$~\cite{pythia}.}
\label{fig:afb_prediction}
\end{figure}

The SM leading order (LO) prediction~\cite{pythia}
for $A_{FB}$ as a function of the dielectron invariant mass ($M_{ee}$) is shown in Fig.~\ref{fig:afb_prediction}
for $u\bar{u}\rightarrow Z/\gamma^*\rightarrow e^+e^-$, $d\bar{d}\rightarrow Z/\gamma^*\rightarrow e^+e^-$, 
and $p\bar{p}\rightarrow Z/\gamma^*\rightarrow e^+e^-$ with the CTEQ6L1 parton distribution functions (PDFs)~\cite{cteq}.
Around the $Z$ pole, the asymmetry is proportional to
both the vector and axial-vector couplings of the $Z$ boson to the
fermions and is numerically close to 0. At large invariant mass, the asymmetry
is dominated by $Z/\gamma^*$ interference and is almost constant
($\approx 0.6$). In the high mass region, the $A_{FB}$ measurement
can be used to investigate possible new phenomena that may alter
$A_{FB}$, such as new neutral gauge bosons or large extra
dimensions~\cite{zprime, led, highmass, highmass_1, highmass_2, 
highmass_3, highmass_4, highmass_5, highmass_CDF}.

In the vicinity of the $Z$ pole, $A_{FB}$ is sensitive to the
effective weak mixing angle ($\stwefffermion$) for each fermion species, 
$f$, involved in a particular measurement. 
To all orders in perturbation theory~\cite{pdg, lep_sinthetaW}, $\stwefffermion$ 
is related to the vector and axial-vector couplings by the expression
\begin{equation}
   g_V^f/g_A^f = 1 - 4|q_f| \stwefffermion .
\end{equation}

This charged lepton effective mixing angle $\stwefflep$ varies as a function of
the momentum transfer at which it is measured. Conventionally,
it is quoted at the $Z$ pole
$\left [\stwefflep(M_Z)\right ]$, and it is identical for 
$e$, $\mu$, and $\tau$ leptons, due to lepton universality. 

In the SM, asymmetries measured at the $Z$ pole~\cite{lep_sinthetaW} depend 
only on the value of $\stwefffermion$ for the fermions 
being considered. Because of the small ratio of vector and axial-vector couplings 
for leptons, the sensitivity of leptonic asymmetries to the changes in effective 
mixing angle arises predominantly through the variation of the leptonic couplings and not 
those of the quarks. 
Therefore, it is customary to express $A_{FB}$ measurements in terms of $\stwefflep$.
In order to extract $\stwefflep$ from $A_{FB}$ under a consistent SM definition and 
compare results with previous measurements, we take into account the difference 
between the electroweak radiative corrections for electrons and $u$/$d$ quarks using 
the relations~\cite{lep_sinthetaW, zgrad, resbos}
\begin{equation}
 \begin{array}{l}
  \displaystyle \stweffu = \stwefflep - 0.0001, \\
  \displaystyle \stweffd = \stwefflep - 0.0002.
 \end{array}
\end{equation}

Precise determinations of $\stwefflep$ have been made in many processes at different $Q^2$ scales. They
include atomic parity violation ($|Q^2| \approx 10^{-18}$~GeV$^2$)~\cite{atomic_PV}, 
M\o%
ller scattering using a polarized electron beam and unpolarized target 
($|Q^2| \approx 0.03$~GeV$^2$)~\cite{Moller}, the NuTeV deep inelastic neutrino and
anti-neutrino scattering on iron ($|Q^2| \approx 4$~GeV$^2$)~\cite{NuTeV}, 
and a number of measurements employing
$e^+e^-$ collisions by the LEP and SLD Collaborations ($|Q^2| \approx
M_Z^2$)~\cite{lep_sinthetaW}. The current world average value of
$\stwefflep$ is 0.23153$\pm$0.00016~\cite{lep_sinthetaW}. 
The two most precise determinations of $\stwefflep$ come from 
the $b$-quark forward-backward asymmetry at LEP,
$A_{FB}^{0,b}$, with $\stwefflep = 0.23221\pm0.00029$, and the left-right
asymmetry at SLD, $A_{lr}(\text{SLD})$, with  
$\stwefflep = 0.23098\pm0.00026$. These two
measurements differ from each other by
about three standard deviations, and deviate by $+$2.1 standard deviations and
$-$1.8 standard deviations from the global fit, respectively.

The LEP Collaborations also measured $\stwefflep$ from the inclusive
hadronic charge asymmetry ($Q_{FB}^{\text{had}}$), with larger uncertainties
governed by the ambiguity of charge separation for final state quark species. 
Furthermore, the hadronic charge asymmetry arising from $u$- and $d$-type  
quarks are in opposite directions, partially canceling.
Thus, modifications to the SM that would affect only $u$ and $d$ quark
couplings are poorly constrained. 
Drell-Yan processes at hadron colliders, in which the initial state is dominated 
by the light $u$ and $d$ quarks in the proton, 
provide a much less ambiguous measurement of the light quark couplings. The
dominant systematic uncertainty at the Tevatron originates from
the quark composition of the proton and anti-proton, which has been well constrained and parametrized by 
PDFs~\cite{cteq}. The use of the Collins-Soper frame 
reduces possible confounding effects from higher order quantum chromodynamics (QCD) corrections.

Previous direct measurements of $u$ and $d$ quark
couplings to the $Z$ boson are of limited precision  
(\cite{lep_sinthetaW, cdf_RunII, H1}). 
With precise determination of the leptonic couplings from LEP and SLD, 
we can interpret the measurement of the forward-backward asymmetry directly 
in terms of the vector and axial-vector couplings of the $u$/$d$ quarks.

At the Tevatron, measurements of the $A_{FB}$, $\stwefflep$, $g_V^{u(d)}$ 
and $g_A^{u(d)}$ have been performed by the CDF and D0
Collaborations~\cite{cdf_RunII, cdf_RunI, d0_RunI, d0_RunII}. The
largest integrated luminosity used for these measurements was 1.1
fb$^{-1}$ for $A_{FB}$ and $\stwefflep$
measurements~\cite{d0_RunII}, and 72 pb$^{-1}$ for $g_V^{u(d)}$ and
$g_A^{u(d)}$ measurements~\cite{cdf_RunII}. In this Article we present
new measurements of the quantities $\stwefflep$, $g_V^{u(d)}$ and $g_A^{u(d)}$ 
based on 5.0 fb$^{-1}$ of integrated luminosity~\cite{d0lumi},
collected using the D0 detector~\cite{d0det} between
April 2002 and April 2009.

\section{Apparatus and Event selection}
\indent The D0 detector~\cite{d0det} comprises a central tracking
system, a calorimeter and a muon system. The central tracking 
system is composed of a silicon microstrip tracker (SMT) and a central
fiber tracker (CFT), both located within a 2~T superconducting
solenoidal magnet and optimized for tracking and vertexing
capabilities at detector pseudorapidities of $|\eta_{\text{det}}|<3$~\cite{d0_coordinate}.
Three liquid argon and uranium calorimeters provide coverage of
$|\eta_{\text{det}}|<3.2$ for electrons with gaps between cryostats 
creating an inefficient electron detection region between $1.0 <|\eta_{\text{det}}| < 1.5$.
The electromagnetic (EM) section of the calorimeter is segmented
into four longitudinal layers with transverse segmentation of
$\Delta \eta \times \Delta \phi = 0.1\times 0.1$, except for the third
layer, where it is $0.05\times 0.05$. The calorimeter is well suited
for a precise measurement of electron and photon energies, providing
a resolution of $\approx$ 3.6\% at an incident energy of $\approx$ 50~GeV. The muon
system surrounds the calorimetry and consists of three layers of
scintillators and drift tubes and 1.8 T iron toroids with a coverage
of $|\eta_{\text{det}}|<2$. 
The three-level trigger system and data acquisition systems are designed to accommodate
the high instantaneous luminosity of Run II. 
A logical OR of dielectron triggers is used to 
collect the data, resulting in a trigger efficiency close to 100\% for signal
events that passed the offline selection.

To select $\zgamma\rightarrow e^+e^-$ events, we require two EM shower candidates with
transverse energy $E_T>25$~GeV measured in the calorimeter. An
isolation cut is imposed on the candidates, requiring that the fraction of
their energy in an annular central (endcap) calorimeter cone of radius $0.2<\Delta
\cal{R}<$ $0.4$ must be less than 15\% (10\%) of the energy in the cone of
$\Delta \cal{R}<$ $0.2$, where $\Delta \cal{R}=$$~\sqrt{(\Delta
\eta)^2 + (\Delta \phi)^2}$. The candidates are further required to
have a significant fraction of their energy deposited in the EM calorimeter
compared to that in the hadron calorimeter, and to have a shower shape
consistent with that expected for an electron. At least one electron
candidate is required to be in the central ($|\eta_{\text{det}}|<1.0$) fiducial
region and spatially matched to a reconstructed
track, while the other candidate may be either in the central or endcap
($1.5<|\eta_{\text{det}}|<2.5$) calorimeter. No track requirement is imposed on
candidates in the endcap calorimeter, since the track reconstruction efficiency is degraded in this region.
If an event has both candidates in the central
calorimeter (CC events), the two candidates are further required to
have opposite charges. For events with one candidate in the central
and the other in the endcap calorimeter (CE events), the 
charge of the central EM candidate is used to determine if it is a forward 
or a backward event. 
To suppress multijet background in CE
events, the electron candidates in the endcap calorimeter are required to 
pass isolation criteria in the tracker,
requiring the scalar sum of the transverse momenta of tracks in the 
annulus $0.05$ $<\Delta \cal{R}<$ $0.4$ centered around the
electron direction to be smaller than 1.5~GeV. 
Events are further required to have the reconstructed $p\bar{p}$ interaction vertex within
40~cm of the detector center in the coordinate along the $z$ axis 
and a reconstructed invariant mass of
the electron pair ($M_{ee}$) between 50 and 1000~GeV.

A total of 157,553 events remain after application of all selection
criteria, with 73,755 CC events and 83,798 CE events. The
forward-backward charge asymmetries are measured in 15 $M_{ee}$ bins
in the range $50<M_{ee}<1000$~GeV. The bin widths are chosen considering the 
statistics of the sample and the mass resolution of the detector. 
The bin widths
and the numbers of forward and backward events for each mass bin are
listed in Table~\ref{Tab:data_num}.

{\begingroup
\begin{table*}
\begin{center}
\caption{Numbers of forward and backward CC and CE events 
in each $M_{ee}$ bin after all selections.}
\begin{tabular}{r@{$\, - \,$}lcccc}
\hline
\hline
\multicolumn{2}{c}{\multirow{2}{*}{$M_{ee}$ (GeV)}}  & \multicolumn{2}{c}{ CC}  &  \multicolumn{2}{c}{ CE} \\ 
               \multicolumn{2}{c}{}  & Forward & Backward & Forward & Backward \\ \cline{1-6}
50 & 60 & 276 & 319 & 54 & 70 \\ 
60 & 70 & 464 & 711 & 238 & 413 \\ 
70 & 75 & 411 & 545 & 285 & 495 \\ 
75 & 81 & 852 & 1062 & 778 & 1240 \\ 
81 & 86.5 & 3359 & 3559 & 3804 & 4245 \\
86.5 & 89.5 & 6681 & 6642 & 8339 & 7591 \\ 
89.5 & 92 & 9297 & 8717 & 11098 & 9534 \\ 
92 & 97 & 12076 & 11109 & 14281 & 11412 \\ 
97 & 105 & 2890 & 2173 & 3711 & 2150 \\ 
105 & 115 & 680 & 431 & 1125 & 395 \\ 
115 & 130 & 408 & 189 & 764 & 229 \\ 
130 & 180 & 439 & 150 & 845 & 269 \\ 
180 & 250 & 138 & 61 & 229 & 73 \\ 
250 & 500 & 63 & 45 & 86 & 24 \\ 
500 & 1000 & 7 & 1 & 1 & 0 \\ \hline 
\hline
\end{tabular}
\label{Tab:data_num}
\end{center}
\end{table*}
\endgroup}

\section{Signal and backgrounds}
Monte Carlo (MC) samples for the $\zgamma \rightarrow e^{+}e^{-}$
process are generated using the {\sc pythia} event generator with
CTEQ6L1 PDFs,
followed by a detailed {\sc geant}-based simulation~\cite{geant} of
the D0 detector response. 
This simulation is then improved by corrections for observed deficiencies in 
the detector simulation and for higher order physical effects not included in {\sc pythia}.

\begin{figure}[htbp]
\epsfig{file=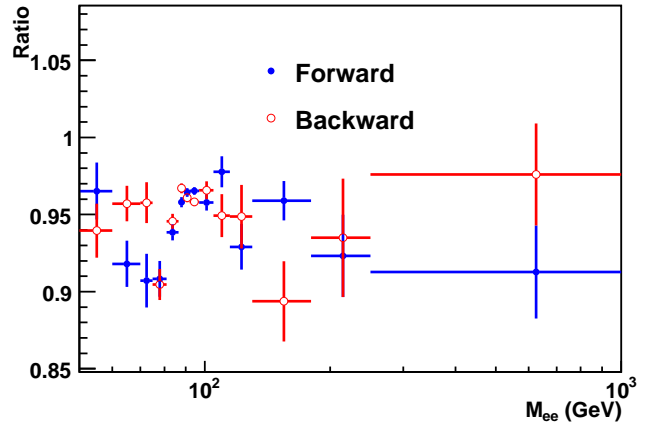, scale = 0.45}
\caption{\small [color online] The data/MC ratio of electron selection efficiencies as a function of
invariant mass, for forward and backward events.}
\label{fig:fb_eff}
\end{figure}

The event selection criteria can result in different 
efficiencies for forward and backward events. 
The electron selection efficiencies are independently measured 
from $Z/\gamma^{*} \rightarrow e^+e^-$ events in data and in the MC, 
where one electron is selected in the central calorimeter using tight 
calorimeter shower shape cuts and track quality cuts, and the second 
electron is used as a probe to determine the detection efficiencies.
These efficiencies are measured for forward and backward events separately. 
For data, the background in each mass bin is estimated and subtracted prior
to the measurement of the efficiencies. 
The ratios between data and MC electron selection efficiencies 
for forward and backward candidates as a function of $M_{ee}$ for electrons 
in the central calorimeter are shown in Fig.~\ref{fig:fb_eff}. 
The ratios are constant within
statistical uncertainties, with the largest deviations observed in a few mass
bins around 70 and 130~GeV. Efficiency corrections derived using data presented on Fig.~\ref{fig:fb_eff} are applied to the MC 
separately for forward and backward events to account for the mis-modeling
of electrons' shower shapes and track matching efficiencies.
In addition, 
the MC is adjusted to
reproduce the calorimeter energy scale and resolution, as well as
the distributions of the instantaneous luminosity and the event
vertex position observed in data.

Next-to-next-to-leading order
(NNLO) QCD corrections~\cite{kfactor} for $\zgamma$ boson production
are applied to the simulated {\sc pythia} sample by reweighting the
$M_{ee}$ distribution, and non-perturbative and next-to-leading order (NLO) corrections 
by reweighting the $\zgamma$ boson transverse momentum and
rapidity distributions~\cite{resbos, NLO_corr}.
The effective mixing angle must be corrected to
include higher order quantum electrodynamics (QED) and weak interaction contributions 
that are not present in our MC.
These higher order corrections are determined 
using the {\sc zgrad2} program~\cite{zgrad}.

The largest background originates from multijet events in which jets are
mis-reconstructed as electrons. Smaller background contributions
arise from other SM processes that produce at least one real
electron or photon in the final state. 
SM backgrounds, such as $Z/\gamma^* \rightarrow \tau\tau$,
$W+X$, $WW$, $WZ$, $\gamma\gamma$, and $t\bar{t}$, are estimated
using the MC. Higher order corrections to the 
cross sections have been applied~\cite{NLO_corr, WW_NLO_corr,
ttbar_NLO_corr}. 
The multijet background is
estimated using collider data by fitting the $M_{ee}$
distribution in the $Z$ pole region (with other SM backgrounds
subtracted) to the sum of the shape predicted by the corrected
$\zgamma \rightarrow e^+e^-$ signal MC and the shape
measured from a multijet-enriched sample. 
The multijet-enriched
sample is selected by reversing the shower shape requirement on the
two electron candidates. The average multijet background fraction
over the entire mass region is found to be approximately $0.9\%$.
The numbers of background events from each source in
forward and backward samples are listed in
Tables~\ref{Tab:all_bkg_qcd_f} and~\ref{Tab:all_bkg_qcd_b}.

\begin{figure}[htbp]
\epsfig{file=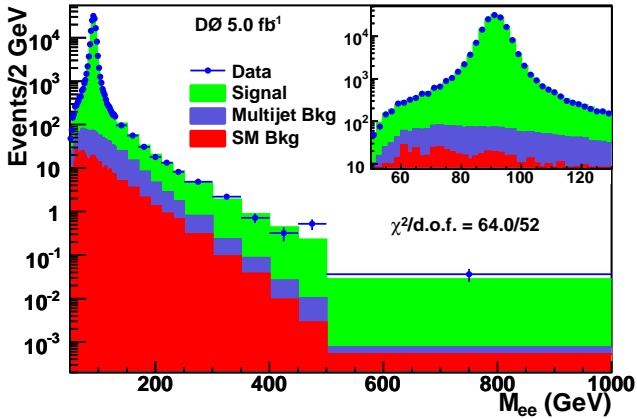,scale=0.45}
\caption{\small [color online] Comparisons of the dielectron invariant mass between data and the sum of 
signal and background predictions for combined CC and CE events. 
The insert focuses on the $Z$ pole region from 50~GeV 
to 130~GeV, where good agreement between data and the sum of signal and 
background predictions is essential to perform the unfolding.}
\label{fig5:invmass}
\end{figure}

\begin{figure}[htbp]
\epsfig{file=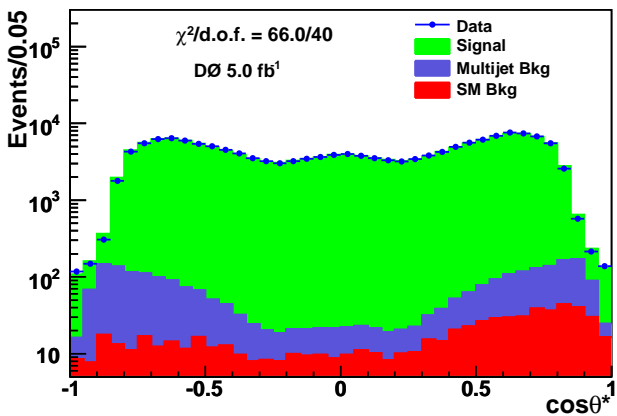,scale=0.45}
\caption{\small [color online] Comparisons of the $\cos \theta^{*}$ between data and the sum of signal and background predictions for combined CC and CE events.} 
\label{fig5:collin}
\end{figure}

Comparisons of data and the sum of signal and background for
$M_{ee}$ and $\cos \theta^*$ are shown in Figs.~\ref{fig5:invmass}
and~\ref{fig5:collin}. 
In the $M_{ee}$ bin 450--500~GeV, the data differ from the SM prediction by 1.8
standard deviations. 
Reasonable agreement is observed for all distributions in both
forward and backward samples for all 15 $M_{ee}$ bins.
The CC and CE raw $A_{FB}$ (not yet unfolded) distributions as functions of $M_{ee}$ are
then calculated from background-subtracted data.

{\begingroup
\begin{table*}
\begin{center}
\caption{Estimated number of background events in each $M_{ee}$ bin in the forward sample.} 
\begin{tabular}{r@{$\, - \,$}lccccccc}
\hline
\hline
\multicolumn{2}{c}{$M_{ee}$ (GeV)} & ~~~~$Z/\gamma^* \rightarrow \tau\tau$~~~~ & ~~~~$W+X$~~~~ & ~~~~$WW$~~~~ & ~~~~$WZ$~~~~ & ~~~~$\gamma\gamma$~~~~ & ~~~~$t\bar{t}$~~~~ & ~~Multijet~~ \\ \hline
 50 & 60     & 12.5 $\pm$ 0.91 &  11.7 $\pm$ 4.41 &  1.65 $\pm$ 0.14 &  0.14 $\pm$ 0.01 &  1.02 $\pm$ 0.35 &  1.00 $\pm$ 0.11 &  38.6 $\pm$ 0.04 \\
 60 & 70     & 29.4 $\pm$ 1.44 &  20.4 $\pm$ 7.17 &  3.29 $\pm$ 0.24 &  0.22 $\pm$ 0.01 &  3.10 $\pm$ 0.43 &  1.83 $\pm$ 0.21 &  105. $\pm$ 0.10 \\
 70 & 75     & 16.6 $\pm$ 0.97 &  17.2 $\pm$ 4.38 &  1.68 $\pm$ 0.14 &  0.18 $\pm$ 0.01 &  1.08 $\pm$ 0.26 &  1.09 $\pm$ 0.13 &  69.2 $\pm$ 0.09 \\
 75 & 81     & 14.5 $\pm$ 0.91 &  16.6 $\pm$ 4.86 &  1.55 $\pm$ 0.13 &  0.31 $\pm$ 0.01 &  1.59 $\pm$ 0.28 &  1.37 $\pm$ 0.15 &  85.8 $\pm$ 0.10 \\
 81 & 86.5   & 5.16 $\pm$ 0.72 &  21.4 $\pm$ 8.33 &  1.82 $\pm$ 0.14 &  0.80 $\pm$ 0.03 &  2.26 $\pm$ 0.31 &  1.16 $\pm$ 0.14 &  80.3 $\pm$ 0.10 \\
 86.5 & 89.5 & 0.94 $\pm$ 0.49 &  8.03 $\pm$ 2.67 &  1.10 $\pm$ 0.12 &  1.72 $\pm$ 0.06 &  1.16 $\pm$ 0.26 &  0.56 $\pm$ 0.07 &  40.6 $\pm$ 0.07 \\
 89.5 & 92   & 1.63 $\pm$ 0.69 &  9.73 $\pm$ 3.07 &  0.86 $\pm$ 0.11 &  2.79 $\pm$ 0.11 &  0.29 $\pm$ 0.24 &  0.68 $\pm$ 0.08 &  31.0 $\pm$ 0.07 \\
 92 & 97     & 1.04 $\pm$ 0.49 &  18.9 $\pm$ 5.34 &  1.64 $\pm$ 0.14 &  3.66 $\pm$ 0.13 &  2.00 $\pm$ 0.30 &  0.99 $\pm$ 0.12 &  62.7 $\pm$ 0.10 \\
 97 & 105    & 1.38 $\pm$ 0.14 &  24.4 $\pm$ 10.8 &  2.72 $\pm$ 0.17 &  0.80 $\pm$ 0.03 &  2.00 $\pm$ 0.31 &  1.50 $\pm$ 0.15 &  88.5 $\pm$ 0.12 \\
 105 & 115   & 1.21 $\pm$ 0.11 &  23.4 $\pm$ 11.3 &  3.03 $\pm$ 0.19 &  0.33 $\pm$ 0.01 &  1.91 $\pm$ 0.30 &  1.22 $\pm$ 0.12 &  94.5 $\pm$ 0.12 \\
 115 & 130   & 1.33 $\pm$ 0.11 &  30.0 $\pm$ 15.5 &  4.33 $\pm$ 0.25 &  0.38 $\pm$ 0.01 &  2.00 $\pm$ 0.29 &  2.16 $\pm$ 0.19 &  108. $\pm$ 0.13 \\
 130 & 180   & 2.38 $\pm$ 0.50 &  41.3 $\pm$ 27.6 &  9.87 $\pm$ 0.55 &  0.82 $\pm$ 0.03 &  4.51 $\pm$ 0.27 &  3.94 $\pm$ 0.40 &  174. $\pm$ 0.17 \\
 180 & 250   & 0.57 $\pm$ 0.03 &  17.1 $\pm$ 14.5 &  4.53 $\pm$ 0.26 &  0.48 $\pm$ 0.02 &  2.84 $\pm$ 0.18 &  1.73 $\pm$ 0.17 &  42.5 $\pm$ 0.08 \\
 250 & 500   & 0.19 $\pm$ 0.02 &  4.80 $\pm$ 3.39 &  1.93 $\pm$ 0.15 &  0.25 $\pm$ 0.01 &  1.24 $\pm$ 0.12 &  0.53 $\pm$ 0.06 &  8.70 $\pm$ 0.04 \\
 500 & 1000  & \ \ \ \ \ \   $<$   0.01 &  \ \ \ \ \ \   $<$   0.01 &  0.07 $\pm$ 0.00 &  \ \ \ \ \ \    $<$   0.01 &  0.04 $\pm$ 0.03 & \ \ \ \ \ \   $<$   0.01 &  0.02 $\pm$ 0.00 \\
\hline
\hline
\end{tabular}
\label{Tab:all_bkg_qcd_f}
\end{center}
\end{table*}
\endgroup}

{\begingroup
\begin{table*}
\begin{center}
\caption{Estimated number of background events in each $M_{ee}$ bin in the backward sample.} 
\begin{tabular}{r@{$\, - \,$}lccccccc}
\hline
\hline
\multicolumn{2}{c}{$M_{ee}$ (GeV)} & ~~~~$Z/\gamma^* \rightarrow \tau\tau$~~~~ & ~~~~$W+X$~~~~ & ~~~~$WW$~~~~ & ~~~~$WZ$~~~~ & ~~~~$\gamma\gamma$~~~~ & ~~~~$t\bar{t}$~~~~ & ~~Multijet~~ \\ \hline
 50 & 60     & 7.52 $\pm$ 0.80 &  4.27 $\pm$ 1.83 &  2.90 $\pm$ 0.20 &  0.20 $\pm$ 0.01 &  0.54 $\pm$ 0.25 &  1.40 $\pm$ 0.16 &  38.8 $\pm$ 0.04 \\
 60 & 70     & 26.2 $\pm$ 1.36 &  4.20 $\pm$ 2.39 &  3.73 $\pm$ 0.24 &  0.28 $\pm$ 0.01 &  2.85 $\pm$ 0.35 &  2.06 $\pm$ 0.24 &  108. $\pm$ 0.10 \\
 70 & 75     & 12.4 $\pm$ 0.85 &  3.06 $\pm$ 1.33 &  2.64 $\pm$ 0.18 &  0.18 $\pm$ 0.01 &  2.32 $\pm$ 0.32 &  1.12 $\pm$ 0.12 &  70.2 $\pm$ 0.09 \\
 75 & 81     & 5.13 $\pm$ 0.72 &  2.74 $\pm$ 1.30 &  2.66 $\pm$ 0.18 &  0.34 $\pm$ 0.01 &  1.66 $\pm$ 0.28 &  1.21 $\pm$ 0.14 &  85.9 $\pm$ 0.10 \\
 81 & 86.5   & 1.10 $\pm$ 0.50 &  5.30 $\pm$ 2.49 &  1.68 $\pm$ 0.13 &  0.76 $\pm$ 0.03 &  2.33 $\pm$ 0.32 &  1.08 $\pm$ 0.13 &  78.8 $\pm$ 0.11 \\
 86.5 & 89.5 & 0.53 $\pm$ 0.49 &  4.25 $\pm$ 1.56 &  1.66 $\pm$ 0.14 &  1.59 $\pm$ 0.06 &  1.44 $\pm$ 0.27 &  0.45 $\pm$ 0.07 &  39.5 $\pm$ 0.08 \\
 89.5 & 92   & 0.15 $\pm$ 0.09 &  3.80 $\pm$ 1.67 &  1.32 $\pm$ 0.13 &  2.43 $\pm$ 0.09 &  1.09 $\pm$ 0.26 &  0.53 $\pm$ 0.07 &  31.5 $\pm$ 0.07 \\
 92 & 97     & 0.24 $\pm$ 0.10 &  2.34 $\pm$ 0.94 &  2.04 $\pm$ 0.14 &  3.17 $\pm$ 0.11 &  2.68 $\pm$ 0.34 &  1.05 $\pm$ 0.12 &  64.1 $\pm$ 0.10 \\
 97 & 105    & 0.30 $\pm$ 0.03 &  6.98 $\pm$ 2.64 &  3.93 $\pm$ 0.24 &  0.75 $\pm$ 0.02 &  2.20 $\pm$ 0.31 &  1.74 $\pm$ 0.20 &  90.0 $\pm$ 0.12 \\
 105 & 115   & 0.26 $\pm$ 0.03 &  5.47 $\pm$ 2.63 &  3.07 $\pm$ 0.19 &  0.35 $\pm$ 0.01 &  2.72 $\pm$ 0.34 &  1.53 $\pm$ 0.15 &  97.3 $\pm$ 0.12 \\
 115 & 130   & 0.46 $\pm$ 0.10 &  6.22 $\pm$ 3.89 &  3.25 $\pm$ 0.20 &  0.37 $\pm$ 0.01 &  2.60 $\pm$ 0.33 &  2.30 $\pm$ 0.24 &  110. $\pm$ 0.13 \\
 130 & 180   & 0.76 $\pm$ 0.49 &  17.0 $\pm$ 10.2 &  5.58 $\pm$ 0.29 &  0.63 $\pm$ 0.03 &  4.97 $\pm$ 0.39 &  4.08 $\pm$ 0.40 &  170. $\pm$ 0.17 \\
 180 & 250   & 0.30 $\pm$ 0.48 &  3.49 $\pm$ 2.92 &  2.42 $\pm$ 0.16 &  0.28 $\pm$ 0.01 &  3.01 $\pm$ 0.18 &  1.70 $\pm$ 0.17 &  46.3 $\pm$ 0.08 \\
 250 & 500   & 0.04 $\pm$ 0.01 &  0.62 $\pm$ 0.85 &  0.58 $\pm$ 0.11 &  0.08 $\pm$ 0.00 &  1.28 $\pm$ 0.20 &  0.50 $\pm$ 0.06 &  8.90 $\pm$ 0.04 \\
 500 & 1000  & \ \ \ \ \ \  $<$   0.01 &  \ \ \ \ \ \  $<$   0.01 & \ \ \ \ \ \   $<$   0.01 & \ \ \ \ \ \   $<$   0.01 &  0.03 $\pm$ 0.00 & \ \ \ \ \ \   $<$   0.01 &  0.02 $\pm$ 0.00 \\
\hline
\hline
\end{tabular}
\label{Tab:all_bkg_qcd_b}
\end{center}
\end{table*}
\endgroup}

\section{Detector resolution and Acceptance}
The finite energy resolution in the determination of the track 
curvature may result in the assignment of events in different bins of invariant mass  
and in changes in the forward/backward classifications. 
These bin migration effects in the raw $A_{FB}$ distribution are corrected 
using an unfolding procedure based on the iterative application 
of the matrix inversion method~\cite{matrix_inversion}, 
as in a previous D0 analysis~\cite{d0_RunII}. The CC and CE
raw $A_{FB}$ distributions are unfolded separately and then
combined. 
We correct for both the wrong classification in terms of dielectron
invariant mass and for the wrong forward/backward assignment by defining 
four detector response matrices. 
The response matrix $R_{ij}^{FF}$ represents the probability that an event
which at the generator level was classified to correspond to forward scattering 
angles and to have $M_{ee}$ in the $j$-th bin, to be reconstructed in the $i$-th 
bin in $M_{ee}$, without any change to the forward/backward assignment. We similarly 
define $R_{ij}^{BB}$ matrix for events which are classified as backward both 
at the generator and at the reconstruction level, and the 
$R_{ij}^{FB}$ and $R_{ij}^{BF}$ matrices for events in which the forward/backward
assignment changes due to detector and reconstruction effects. Tests of the unfolding
procedure are performed comparing the generator level distribution with the one
obtained after unfolding the events processed through the full detector simulation
and reconstructed as data.

The bin purity is defined as
the ratio between the number of events generated in a mass bin and also
reconstructed in the same mass bin ($N_{gen}^{\text{reco}}$) and the number of
events reconstructed in this mass bin ($N^{\text{reco}}$). 
The lowest purity occurs for the two mass bins below the $Z$ pole
($81<M_{ee}<86.5$~GeV and $86.5<M_{ee}<89.5$~GeV) and is about 25\%.
Since the corrected MC can describe the data mass spectra 
with finer binning as shown in Fig.~\ref{fig5:invmass}, 
these low purity bins can be well modeled. 
The rest of the mass bins have purity 
varying between 50\% and 96\%.

After unfolding for detector resolution effects, the data are further
corrected for acceptance. Using the corrected signal MC,
we derive corrections for kinematic and geometric acceptance and
for electron identification efficiencies.

\section{Charge misidentification rate}
The electron charge determines if an event is forward
or backward. Mismeasurement of the sign may result in a dilution of
$A_{FB}$. The charge misidentification probability $f_q$ is given by
\begin{equation}
  f_q = \frac{1}{2}N_{SS}/(N_{SS}+N_{OS}),
\end{equation}
where $N_{SS}$ ($N_{OS}$) is the total number of $\zgamma
\rightarrow e^+e^-$ events reconstructed with same-sign
(opposite-sign) electrons. Since few same-sign
events are observed in data, 
the corrected $\zgamma$ MC is used to determine the shape of the misidentification probability as a function of dielectron invariant mass. The overall normalization 
is set by the misidentification probability determined from data around the $Z$ pole. The misidentification probability
is a function of $M_{ee}$ and rises from 0.75\% at
$M_{ee}=50$~GeV to 3.2\% for $M_{ee}>500$~GeV. The charge
misidentification probability is included as a dilution factor $\cal{D}$
in $A_{FB}$, with ${\cal{D}}=(1-2f_q)/(1-2f_q+f^2_q)$ for CC events
and ${\cal{D}}=(1-2f_q)$ for CE events.

\section{Systematic uncertainties}
The systematic uncertainties in the measurement of $\stwefflep$,
and of the unfolded distribution of $A_{FB}$, $g_V$, and $g_A$, are listed below.

\begin{itemize}
\item{\bf PDF}\\ 
Uncertainties in the input parton distributions lead to uncertainties in the event acceptance. The systematic uncertainty
 due to the PDFs uncertainty is estimated
by reweighting the central PDFs using the 40 CTEQ6.1M error sets, and the
90\% C.L. uncertainty is calculated using the prescription suggested by the
CTEQ group~\cite{cteq}.

\item{\bf Electron energy scale and resolution}\\ 
The energy scale and resolution for electrons in MC are corrected
to match the observed $Z$ boson pole position and width.
The statistical uncertainties of the calibration parameters applied to MC
are considered as a source systematic uncertainties. 
We vary each parameter by $\pm1$ standard deviation to estimate the uncertainty on the final measured quantities.

\item{\bf MC statistics}\\
To determine the systematic uncertainty due to the limited number of MC events,
we divide the MC samples into ten independent sub-samples and perform ten pseudo-experiments.
The spread of the unfolded $A_{FB}$ and measured $\stwefflep$ from 
these pseudo-measurements divided by $\sqrt{10}$ is
assigned as the systematic uncertainty due to the limited MC statistics.

\item{\bf Electron identification efficiency}\\ 
To ensure the MC correctly models the 
electron and event selection efficiencies observed in data, we
apply data/MC efficiency scale factors to the MC for
forward and backward events separately. The bin-by-bin statistical
fluctuations of these correction factors as a function of $M_{ee}$ are
taken into account and are propagated to the systematic
uncertainties.

\item{\bf Background modeling}\\ To estimate the uncertainty due to the multijet background,
we vary the reversed electron shower shape requirements to obtain
different mass spectra of multijet control samples. 
The uncertainties on SM backgrounds estimated using the
MC mainly come from the uncertainties of the
energy smearing, data and MC efficiency scale
factors, and the uncertainty of the inclusive cross section for
each process. For the $W+X$ inclusive background, additional
uncertainties due to the modeling of the electron misidentification probability contributed
by extra jets and the modeling of the $W$ boson $p_T$ (obtained from a comparison of 
{\sc pythia} and {\sc alpgen}~\cite{alpgen_d0}) are also taken into account.

\item{\bf Charge misidentification}\\
The statistical fluctuations in the misidentification probability measured from data in each
mass bin are included as a systematic uncertainty. 

\end{itemize}

The systematic uncertainties on the $\stwefflep$ extraction are summarized in
Table~\ref{Tab:stw_syst}.
The primary systematic uncertainties are due to the PDFs (\stwsysterrpdf)
and the EM energy calibration and resolution (\stwsysterrscale).
A correction factor is introduced to account for higher order electroweak
corrections which are not included in {\sc pythia}. It is determined by
generating {\sc zgrad2} and {\sc pythia} samples and comparing the 
$A_{FB}$ distributions at the generator level. We find that there is a constant
$+0.0005$ positive shift in the full $\stwefflep$ prediction from 
{\sc zgrad2} relative to the LO prediction from {\sc pythia}. 
We add this correction factor to the extracted value of 
$\stwefflep$.

The systematic uncertainties in the unfolded $A_{FB}$ distribution 
are listed in Table~\ref{Tab:afb_syst}.
In addition to the common sources listed above, uncertainties from
higher order corrections and different SM inputs are taken into
consideration. Higher order QCD, QED, and electroweak corrections
can change the $A_{FB}$ predictions and thus induce
additional uncertainty. We compare {\sc pythia} $A_{FB}$ distributions to
those of {\sc zgrad2}~\cite{zgrad} with the $Z/\gamma^*$ boson $p_T$ tuned to
the {\sc resbos}~\cite{resbos} prediction. 
{\sc resbos} has the advantage of including
most of the electroweak effects with a full simulation of the non-perturbative 
and next-to-leading-logarithm (NLL) QCD effects.
The difference between the two predictions is taken as
a systematic uncertainty.
Different input values of $\stwefflep$ in {\sc pythia} will change the 
kinematic and geometric acceptances, and thus introduce uncertainty into
the unfolding assumptions. We take $\stwefflep=0.232$ as the default input value, 
and vary it by the measured $\stwefflep$ uncertainty
(0.001). We then repeat the unfolding procedure and take the largest
deviation from the unfolded $A_{FB}$ as the uncertainty.

The $g_V^q$ and $g_A^q$ couplings are extracted from the unfolded
$A_{FB}$ distribution and thus include all of the uncertainties 
(statistical and systematic) that affect the $A_{FB}$ 
measurement. Additional uncertainties on the couplings from
predictions with different PDF sets will be discussed later.

\begin{table*}
\begin{center}
\caption{Uncertainties for the $\stwefflep$ measurement. All uncertainties are symmetric.}
\begin{tabular}{l|l}
\hline
\hline
Uncertainty source&  $\Delta \stwefflep$  \\ \hline
Statistical   &   0.00080     \\ \hline
Systematics   &   0.00061     \\
 \ \ PDF/Acceptance &  0.00048  \\
 \ \ EM scale/resolution &  0.00029 \\
 \ \ MC Statistics &  0.00020 \\
 \ \ Electron identification &  0.00008\\
 \ \ Bkg. modeling &  0.00008 \\
 \ \ Charge misidentification &  0.00004 \\
 \ \ Higher order &  0.00008 \\ \hline
Total uncertainty  &  0.00102 \\
\hline
\hline
\end{tabular}
\label{Tab:stw_syst}
\end{center}
\end{table*}

\begin{table*}
\begin{center}
\caption{Systematic uncertainties per bin for the unfolded $A_{FB}$ measurement. All uncertainties are symmetric.}
\begin{tabular}{r@{$\, - \,$}lcccccccc|c}
\hline
\hline
\multicolumn{2}{c}{$M_{ee}$ (GeV)} & EM scale/ & Electron & Bkg. & MC     & PDF/ & Charge mis-& QCD, & Input  & Total\\
\multicolumn{2}{c}{}               & resolution& identification &  modeling & Statistics   & Acceptance & identification & QED  & $\stwefflep$ & \\ \hline
  50 & 60      &     0.009 &     0.001 &     0.002  &     0.011 &     0.015 &     0.002 &     0.008  &     0.002 &     0.023 \\
  60 & 70      &     0.002 &     0.004 &     0.001  &     0.014 &     0.005 &     0.002 &     0.012  &     0.001 &     0.019 \\
  70 & 75      &     0.003 &     0.003 &     0.001  &     0.015 &     0.003 &     0.002 &     0.012  &    $< 0.001$\ \ \ \  &     0.020 \\
  75 & 81      &     0.002 &     0.001 &    $< 0.001$\ \ \ \   &     0.008 &    $< 0.001$\ \ \ \  &     0.001 &     0.010  &     0.001 &     0.013 \\
  81 & 86.5    &     0.002 &    $< 0.001$\ \ \ \  &    $< 0.001$\ \ \ \   &     0.005 &     0.008 &    $< 0.001$\ \ \ \  &     0.005  &     0.001 &     0.011 \\
  86.5 & 89.5  &     0.001 &    $< 0.001$\ \ \ \  &    $< 0.001$\ \ \ \   &     0.003 &     0.002 &    $< 0.001$\ \ \ \  &     0.002  &     0.001 &     0.004 \\
  89.5 & 92    &     0.001 &    $< 0.001$\ \ \ \  &    $< 0.001$\ \ \ \   &     0.002 &     0.002 &    $< 0.001$\ \ \ \  &     0.002  &     0.001 &     0.003 \\
  92 & 97      &     0.001 &    $< 0.001$\ \ \ \  &    $< 0.001$\ \ \ \   &     0.002 &     0.002 &    $< 0.001$\ \ \ \  &     0.004  &     0.001 &     0.005 \\
  97 & 105     &     0.002 &    $< 0.001$\ \ \ \  &    $< 0.001$\ \ \ \   &     0.005 &     0.007 &     0.001 &     0.008  &     0.001 &     0.012 \\
  105 & 115    &     0.002 &     0.001 &    $< 0.001$\ \ \ \   &     0.008 &     0.002 &     0.002 &     0.011  &     0.001 &     0.014 \\
  115 & 130    &     0.003 &     0.002 &     0.003  &     0.010 &     0.004 &     0.004 &     0.011  &    $< 0.001$\ \ \ \  &     0.017 \\
  130 & 180    &     0.001 &     0.005 &     0.002  &     0.006 &     0.014 &     0.011 &     0.013  &     0.001 &     0.024 \\
  180 & 250    &     0.002 &     0.002 &     0.002  &     0.003 &     0.010 &     0.011 &     0.006  &    $< 0.001$\ \ \ \  &     0.017 \\
  250 & 500    &     0.001 &     0.001 &     0.002  &     0.008 &     0.001 &     0.019 &     0.004  &    $< 0.001$\ \ \ \  &     0.022 \\
  500 & 1000   &     0.001 &    $< 0.001$\ \ \ \  &    $< 0.001$\ \ \ \   &    $< 0.001$\ \ \ \  &    $< 0.001$\ \ \ \  &     0.016 &    $< 0.001$\ \ \ \   &    $< 0.001$\ \ \ \  &     0.016 \\ \hline
\hline
\end{tabular}
\label{Tab:afb_syst}
\end{center}
\end{table*}

\section{measurement of $\boldsymbol{\stwefflep}$}
\begin{figure}[htbp]
\epsfig{file=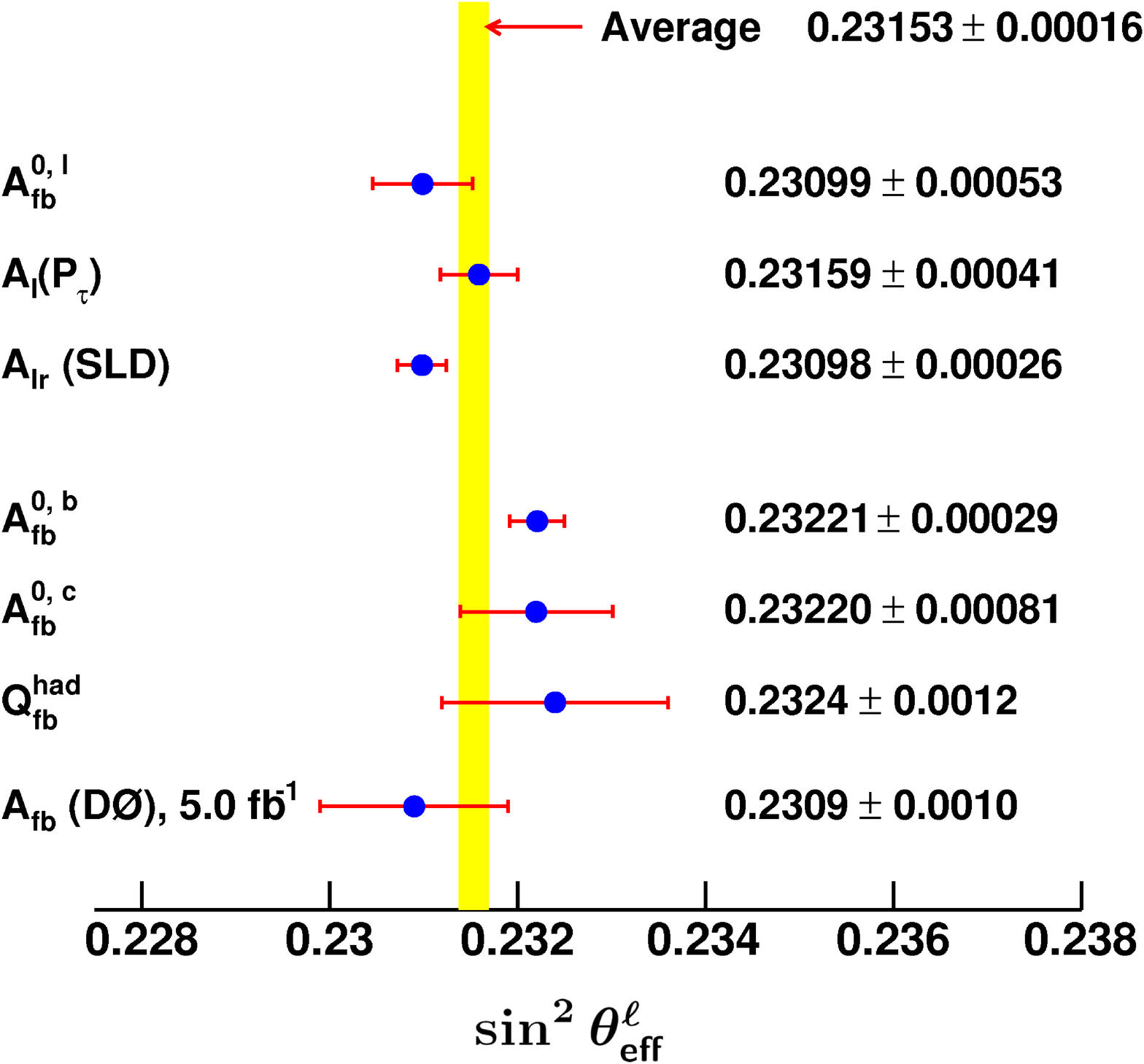,scale = 0.11}
\caption{\small Comparison of measured $\stwefflep$ with results from other experiments. 
The average is a combination of $A_{FB}^{0,\ell}$, $A_{l}(P_{\tau})$, 
$A_{lr}(\text{SLD})$, $A_{FB}^{0,b}$, $A_{FB}^{0,c}$, and $Q_{FB}^{\text{had}}$ measurements from the LEP and 
SLD Collaborations.}
\label{fig:stw_combined}
\end{figure}

The value of $\stwefflep$ is extracted from data by comparing the background-subtracted
raw $A_{FB}$ distribution with simulated $A_{FB}$ templates corresponding to different
input values of $\stwefflep$. 
This procedure avoids the increase of the systematic uncertainty of the measurement 
introduced by the use of the unfolding procedure and maximizes the statistical
significance of the final result. 
Although variations in $\stwefflep$ have some effect over the full mass range $50<M_{ee}<1000$~GeV, 
the central value is
predominantly determined by the events in the $Z$ pole region, where the statistics are highest and 
the effects of background are smallest.
Using events in the range $70<M_{ee}<130$~GeV, 
we measure $\stwefflep =\stwcenter \pm \stwstaterr ~(\mbox{stat.}) \pm \stwsysterr~(\mbox{syst.})$ 
using {\sc pythia}.
We then include higher order electroweak corrections using the {\sc zgrad2} program.
Taking into account the effect of higher order corrections results in a central value of
$\stwefflep=0.2309 \pm \stwstaterr ~(\mbox{stat.}) \pm \stwsysterr~(\mbox{syst.})$.
We also check the $\stwefflep$ predictions using {\sc zgrad2} and
{\sc zfitter}~\cite{zfitter} using the same input SM parameters 
and find the two results are consistent. 
Higher order electroweak and QCD corrections 
included in {\sc zfitter} and not implemented in {\sc zgrad2} have 
a negligible impact on the $\stwefflep$ measurement. 
Therefore, our measured $\stwefflep$ can be directly compared with
the values measured by the LEP and SLD Collaborations~\cite{lep_sinthetaW}.
The comparison is shown in Fig.~\ref{fig:stw_combined}.
The most precise measurements are 
the LEP $b$-quark forward-backward asymmetry, $A_{FB}^{0,b}$, 
the SLD left-right asymmetry, $A_{lr}(\text{SLD})$, 
the LEP $\tau$-lepton polarisation measurement, $A_{l}(P_{\tau})$,
and the SLD lepton asymmetry, $A_{FB}^{0,\ell}$.
Our result is more precise than the LEP combined
inclusive hadronic charge asymmetry measurement,
and comparable in precision with the LEP $c$-quark forward-backward 
asymmetry $A_{FB}^{0,c}$.

\section{Measurement of the unfolded $\boldsymbol{A_{FB}}$ distribution}
\begin{center}
\begin{figure}[htbp]
\epsfig{file=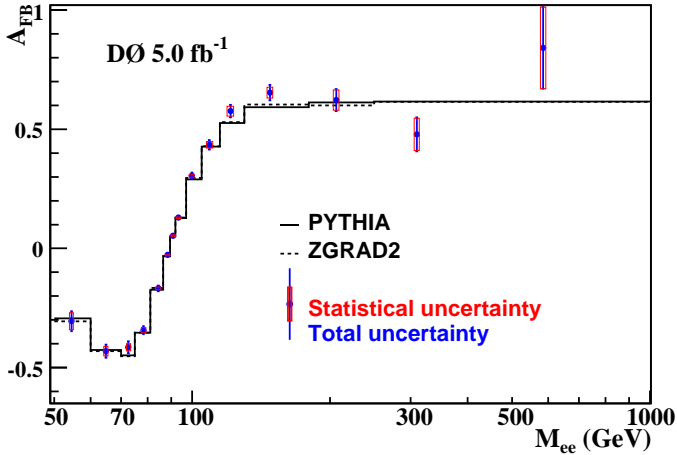, scale=0.45}
\caption{\small Comparison between the unfolded $A_{FB}$ (points) and
the {\sc pythia} (solid curve) and {\sc zgrad2} (dashed line)
predictions. The boxes and vertical lines
show the statistical and total uncertainties, respectively.} 
\label{fig:compare_afb}
\end{figure}
\end{center}

The final unfolded $A_{FB}$ distribution using both CC and CE events
is shown in Fig.~\ref{fig:compare_afb} and Table~\ref{Tab:afb_final}, 
together with {\sc pythia} and {\sc zgrad2} predictions. 
Because of the migration between
mass bins, the correlation matrix is
important for events near the $Z$ pole region. The correlation coefficients
are shown in Table~\ref{Tab:Correlation_All}.
In the mass bins 130--180 and 250--500~GeV 
small deviations ($<2$ standard deviations) are observed. 
The $\chi^2/${d.o.f} between data and prediction is $15.3/15$ for {\sc pythia},
and $12.8/15$ for {\sc zgrad2}.

\begin{table*}
\begin{center}
\caption{The unfolded $A_{FB}$ distribution compared with the theoretical predictions. The first 
column shows the mass ranges used. The second column shows the
cross section weighted average of the invariant mass in each mass bin derived from {\sc pythia}. The
third and fourth columns show the $A_{FB}$ predictions from {\sc pythia} and {\sc zgrad2}.
The last column is the unfolded $A_{FB}$, where the
first uncertainty is statistical and the second is systematic.}
\begin{tabular}{r@{$\, - \,$}lcccc} \hline
\hline
\multicolumn{2}{c}{\multirow{2}{*}{$M_{ee}$ (GeV)}} & \multirow{2}{*}{$\langle M_{ee} \rangle$ (GeV)}  & \multicolumn{2}{c}{\ \ Predicted $A_{FB}$}  &  \ \ \ \multirow{2}{*}{Unfolded $A_{FB}$} \\
         \multicolumn{2}{c}{}      &           & {\ \ \sc pythia} & {\sc zgrad2} &  \\  \hline
50 & 60 & \ 54.5   & $-0.293$ & $-0.307$  &  \ \ \ $ -0.305 \pm 0.036\pm 0.023 $ \\
60 & 70 & \ 64.9   & $-0.426$ & $-0.431$  &  \ \ \ $ -0.431 \pm 0.020\pm 0.019 $ \\
70 & 75 & \ 72.6   & $-0.449$ & $-0.452$  &  \ \ \ $ -0.415 \pm 0.015\pm 0.020 $ \\
75 & 81 & \ 78.3   & $-0.354$ & $-0.354$  &  \ \ \ $ -0.343 \pm 0.011\pm 0.013 $ \\
81 & 86.5 & \ 84.4   & $-0.174$ & $-0.166$  &  \ \ \ $ -0.168 \pm 0.006\pm 0.011 $ \\
86.5 & 89.5 & \ 88.4   & $-0.033$ & $-0.031$  &  \ \ \ $ -0.028 \pm 0.003\pm 0.004 $ \\
89.5 & 92 & \ 90.9   & \ \ $0.051$ & \ \ $0.052$  & \ \ \ \ \ $ 0.054 \pm 0.003\pm 0.003 $ \\
92 & 97 & \ 93.4   & \ \ $0.127$ & \ \ $0.129$  & \ \ \ \ \ $ 0.129 \pm 0.003\pm 0.005 $ \\
97 & 105 & \ 99.9   & \ \ $0.289$ & \ \ $0.296$  & \ \ \ \ \ $ 0.305 \pm 0.007\pm 0.012 $ \\
105 & 115 & 109.1  & \ \ $0.427$ & \ \ $0.429$  & \ \ \ \ \ $ 0.435 \pm 0.014\pm 0.014 $ \\
115 & 130 & 121.3  & \ \ $0.526$ & \ \ $0.530$  & \ \ \ \ \ $ 0.576 \pm 0.021\pm 0.017 $ \\
130 & 180 & 147.9  & \ \ $0.593$ & \ \ $0.603$  & \ \ \ \ \ $ 0.654 \pm 0.022\pm 0.024 $ \\
180 & 250 & 206.4  & \ \ $0.613$ & \ \ $0.600$  & \ \ \ \ \ $ 0.623 \pm 0.043\pm 0.017 $ \\
250 & 500 & 310.5  & \ \ $0.616$ & \ \ $0.615$  & \ \ \ \ \ $ 0.479 \pm 0.068\pm 0.022 $ \\
500 & 1000 & 584.4  & \ \ $0.616$ & \ \ $0.615$  & \ \ \ \ \ $ 0.842 \pm 0.171\pm 0.016 $ \\ \hline
\hline
\end{tabular}
\label{Tab:afb_final}
\end{center}
\end{table*}

\begin{table*}[!htb]
\begin{center}
\caption{Correlation coefficients between different $M_{ee}$ bins.
Only half of the symmetric correlation matrix is presented.}
\begin{tabular}{c|ccccccccccccccc}
\hline
\hline
Mass bin & 1    & 2    & 3    & 4    & 5    & 6    & 7    & 8    & 9    & 10   & 11   & 12   & 13   & 14   & 15   \\ \hline
       1 & 1.00 & 0.24 & 0.03 & 0.00 & 0.00 & 0.00 & 0.00 & 0.00 & 0.00 & 0.00 & 0.00 & 0.00 & 0.00 & 0.00 & 0.00 \\
       2 &      & 1.00 & 0.39 & 0.06 & 0.02 & 0.02 & 0.02 & 0.01 & 0.00 & 0.00 & 0.00 & 0.00 & 0.00 & 0.00 & 0.00 \\
       3 &      &      & 1.00 & 0.46 & 0.11 & 0.04 & 0.03 & 0.02 & 0.00 & 0.00 & 0.00 & 0.00 & 0.00 & 0.00 & 0.00 \\
       4 &      &      &      & 1.00 & 0.51 & 0.15 & 0.07 & 0.04 & 0.01 & 0.00 & 0.00 & 0.00 & 0.00 & 0.00 & 0.00 \\
       5 &      &      &      &      & 1.00 & 0.72 & 0.32 & 0.11 & 0.01 & 0.00 & 0.00 & 0.00 & 0.00 & 0.00 & 0.00 \\
       6 &      &      &      &      &      & 1.00 & 0.78 & 0.40 & 0.03 & 0.00 & 0.00 & 0.00 & 0.00 & 0.00 & 0.00 \\
       7 &      &      &      &      &      &      & 1.00 & 0.80 & 0.13 & 0.01 & 0.00 & 0.00 & 0.00 & 0.00 & 0.00 \\
       8 &      &      &      &      &      &      &      & 1.00 & 0.48 & 0.04 & 0.00 & 0.00 & 0.00 & 0.00 & 0.00 \\
       9 &      &      &      &      &      &      &      &      & 1.00 & 0.38 & 0.03 & 0.00 & 0.00 & 0.00 & 0.00 \\
      10 &      &      &      &      &      &      &      &      &      & 1.00 & 0.28 & 0.01 & 0.00 & 0.00 & 0.00 \\
      11 &      &      &      &      &      &      &      &      &      &      & 1.00 & 0.16 & 0.00 & 0.00 & 0.00 \\
      12 &      &      &      &      &      &      &      &      &      &      &      & 1.00 & 0.06 & 0.00 & 0.00 \\
      13 &      &      &      &      &      &      &      &      &      &      &      &      & 1.00 & 0.06 & 0.00 \\
      14 &      &      &      &      &      &      &      &      &      &      &      &      &      & 1.00 & 0.02 \\
      15 &      &      &      &      &      &      &      &      &      &      &      &      &      &      & 1.00 \\
\hline
\hline
\end{tabular}
\label{Tab:Correlation_All}
\end{center}
\end{table*}

\section{measurement of $\boldsymbol{g_V^{u(d)}}$ and $\boldsymbol{g_A^{u(d)}}$ from the unfolded distribution}
\begin{center}
\begin{figure*}[htbp]
\epsfig{file=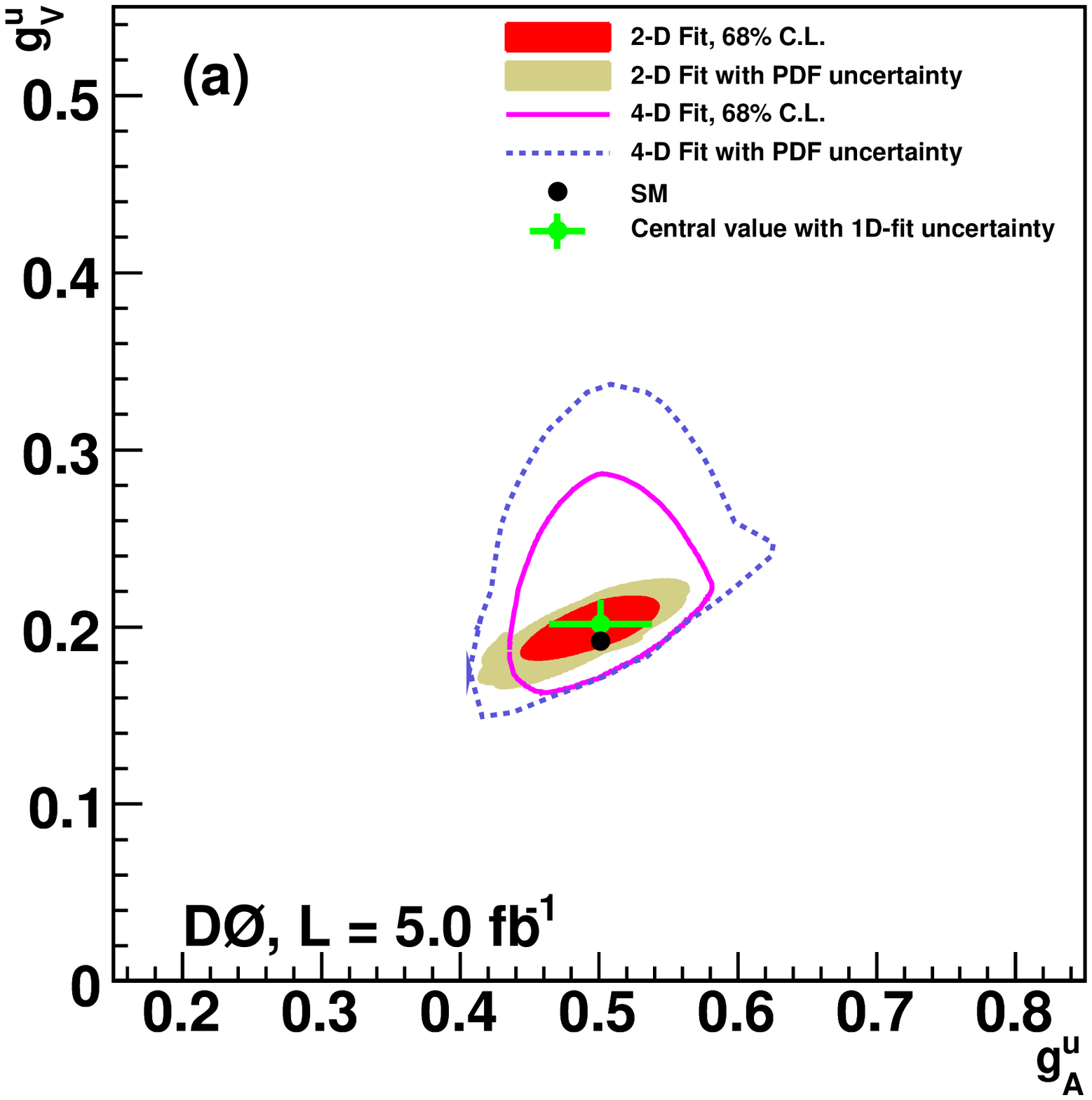, scale=0.42}
\epsfig{file=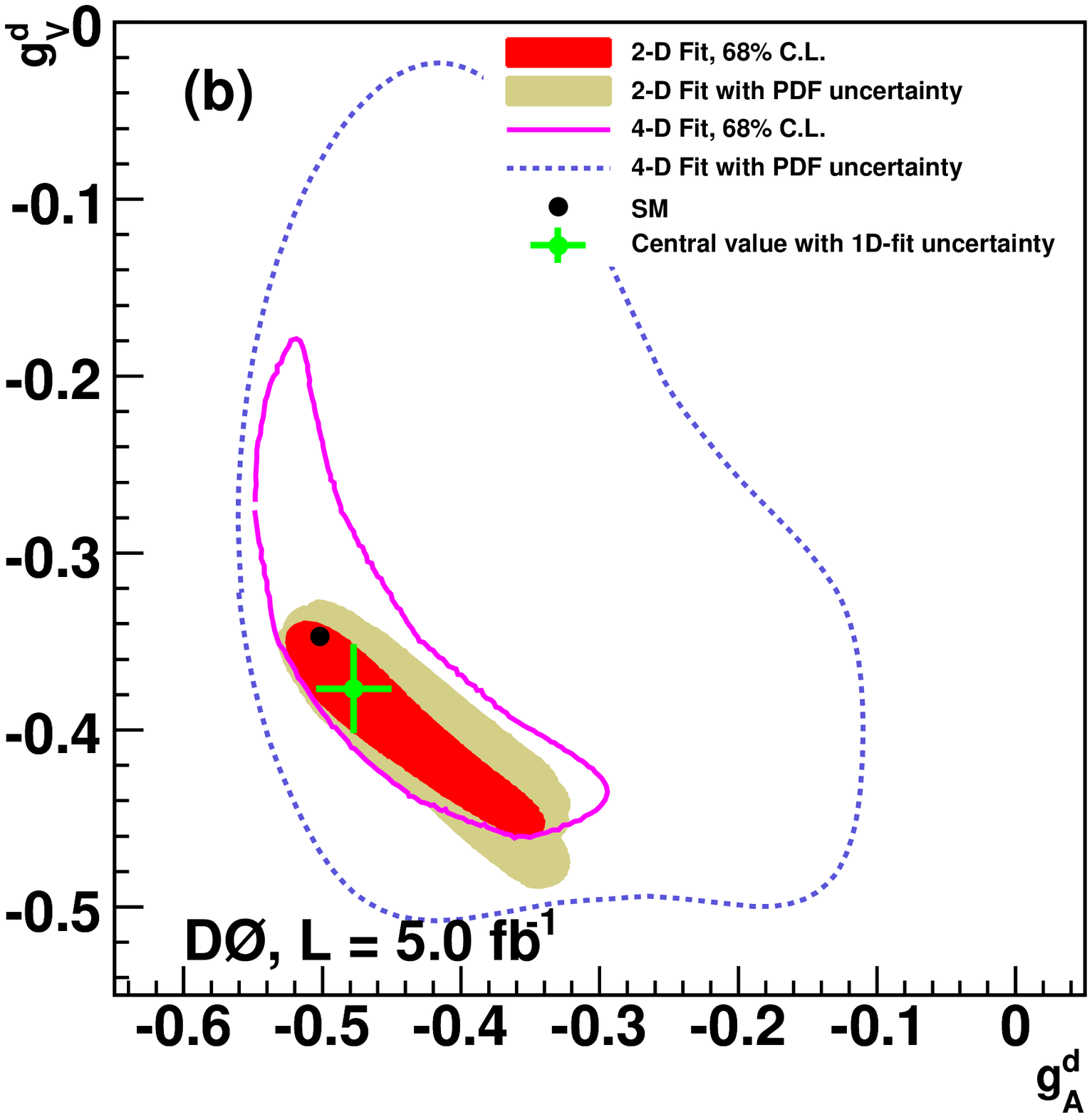, scale=0.42}
\caption{\small
[color online] The 68\% C.L. contours of (a) $g_V^{u}$ and
$g_A^{u}$, and (b) $g_V^{d}$ and $g_A^{d}$ from a two-dimensional and a four-dimensional $\chi^2$ fit 
with statistical and systematic uncertainties. The outer regions are determined by the theoretical PDF uncertainty.
The two-dimensional correlation contours correspond to 
$\Delta \chi^2 = 2.3$ for different $g_A$ and $g_V$ parameters, as obtained
from two-parameter (shaded regions) and four-parameter (solid and dashed curves) fits. The value 2.3
corresponds to the 68\% C.L. region in two dimensions. In the case of four-parameter fit,
the curve is a projection onto the two-dimensional plane of the envelope of the
four-dimensional $\Delta \chi^2 = 4.72$ surface. 
The crosses indicate the best two-dimensional fit values, 
and the uncertainties correspond to the one-dimensional limits.}
\label{fig:zquark}
\end{figure*}
\end{center}

\begin{center}
\begin{figure*}[htbp]
\epsfig{file=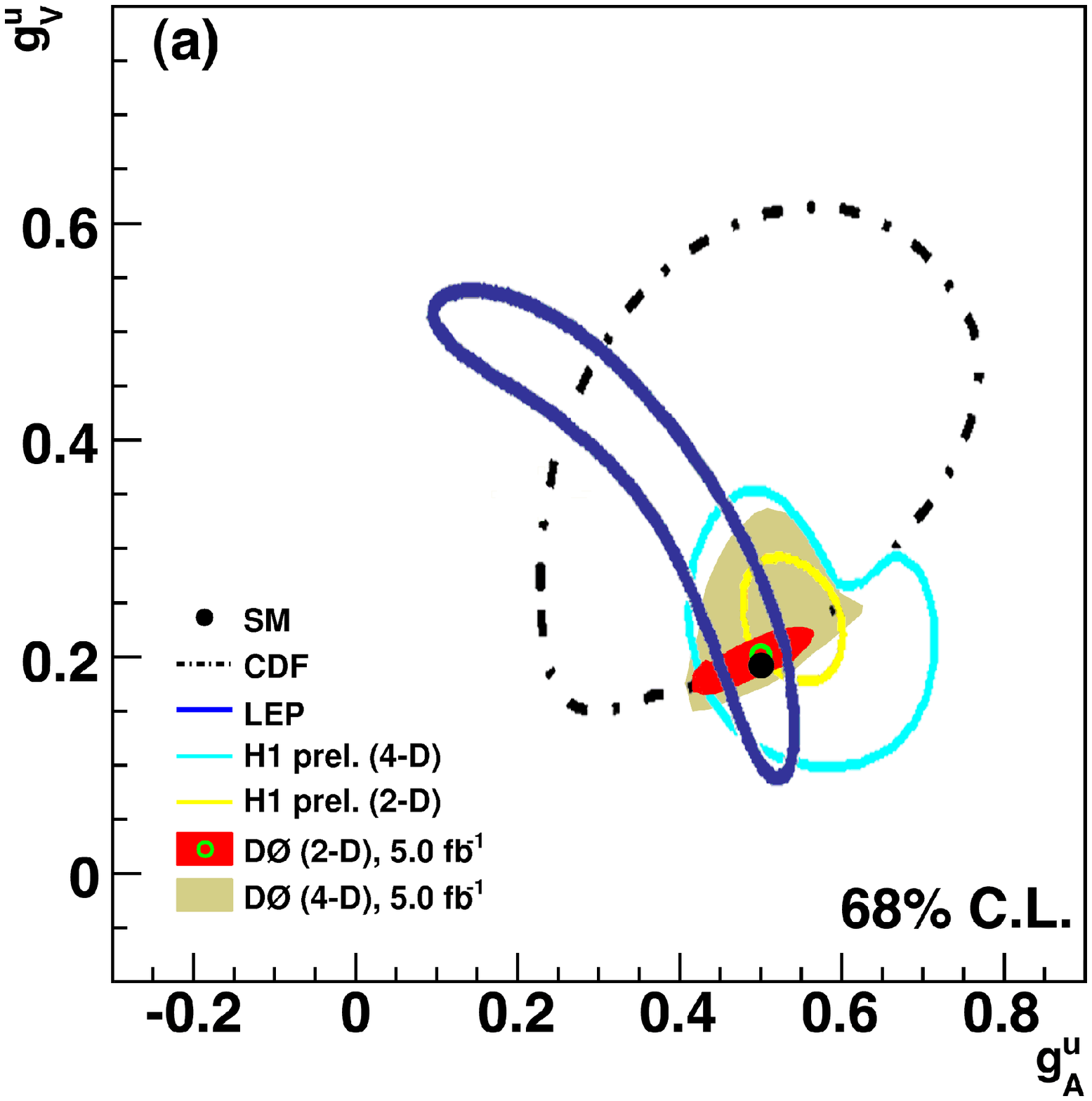, scale=0.42}
\epsfig{file=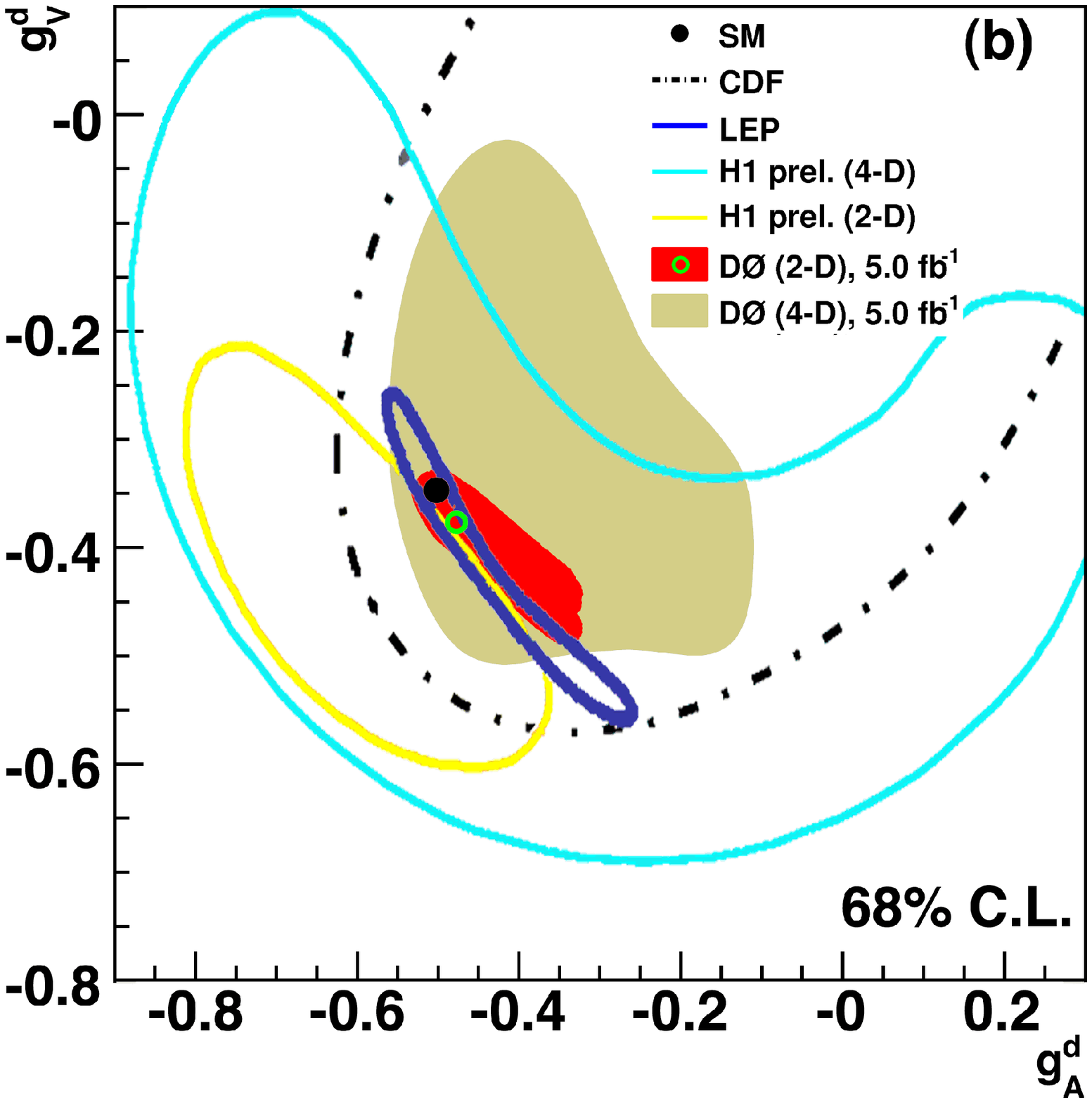, scale=0.42}
\caption{\small
[color online] The 68\% C.L. contours of (a) $g_V^{u}$ and
$g_A^{u}$, and (b) $g_V^{d}$ and $g_A^{d}$ measured by D0 compared with other experiments. 
The LEP and CDF Collaborations performed fits with four free parameters
to determine these couplings, 
while we and the H1 Collaboration performed both two and four free parameters fits.
The LEP results have another solution (not shown) which is excluded by the H1, CDF and D0 results.}
\label{fig:zquark_other}
\end{figure*}
\end{center}

We extract the individual quark couplings by comparing the unfolded $A_{FB}$ distribution to
templates generated with {\sc resbos} for different values of the $Z$-light quark couplings. 
To determine $g_{V}^{u(d)}$ and $g_{A}^{u(d)}$, the couplings of
electrons to $Z$ bosons are fixed to their SM values and
$\stwefflep$ is fixed to the global fit value
0.23153~\cite{lep_sinthetaW}. A two-dimensional $\chi^2$ fit~\cite{minuitfit} is used
to constraint the couplings, and a four-dimensional fit is presented as reference.
The two-dimensional fit is performed by fixing the $u$ quark ($d$ quark) couplings 
to their SM values when fitting $d$ quark ($u$ quark) couplings,
while the four-dimensional fit is performed by letting the $u$ quark and $d$ quark couplings vary simultaneously. 
The best fit values, 
together with results from other experiments, are shown in
Table~\ref{Tab:zquark}. Figure~\ref{fig:zquark} depicts the 68\% C.L.
contours of the $\chi^2$ fit and the contours of the theoretical uncertainty 
from the PDF uncertainties determined using the CTEQ prescription~\cite{cteq}. 
The correlation coefficients between $g_V^{u}$, $g_A^u$, $g_V^d$, and
$g_A^d$ are shown in Table~\ref{Tab:coefficient}, without the PDF uncertainty included. 
The comparisons between different measurements from 
LEP~\cite{lep_sinthetaW}, H1~\cite{H1_2010}, CDF~\cite{cdf_RunII}, 
and D0 are shown in Fig.~\ref{fig:zquark_other}.
Because of the high statistics of our data sample, and the 
reduced ambiguity in the quark content of the initial state, 
these are the world's most precise direct measurements of
$g_V^{u}$, $g_A^{u}$, $g_V^{d}$, and $g_A^{d}$ to date.

\begin{table*}
\begin{center}
\caption{Measured $g_V^{u(d)}$ and $g_A^{u(d)}$ values from
different experiments compared with the SM predictions. The D0
results are derived from best two-dimensional and four-dimensional $\chi^2$ fit, 
given with their total uncertainty. } 
\begin{tabular}{lcccc}
\hline
\hline
 & $g_{A}^u$ & $g_{V}^u$ & $g_{A}^d$ & $g_{V}^d$ \\ \hline
 D0 (2-D) & $0.501 \pm 0.061$ & $0.202 \pm 0.025 $ & $-0.477 \pm 0.112 $  & $-0.377 \pm 0.081$ \\
 D0 (4-D) & $0.501 \pm 0.110$ & $0.201 \pm 0.112 $ & $-0.497 \pm 0.165 $  & $-0.351 \pm 0.251$ \\
CDF~\cite{cdf_RunII}(4-D) & $0.441^{+0.218}_{-0.186}$ & $0.399^{+0.166}_{-0.199}$ & $-0.016^{+0.358}_{-0.544}$  & $-0.226^{+0.641}_{-0.304}$ \\
H1~\cite{H1}(4-D) & $0.56 \pm 0.10$ & $0.05 \pm 0.19$ & $-0.77 \pm 0.37$  & $-0.50 \pm 0.37$ \\
LEP~\cite{lep_sinthetaW}(4-D) & $0.47^{+0.05}_{-0.33}$ & $0.24^{+0.28}_{-0.11}$ & $-0.52^{+0.05}_{-0.03}$  & $-0.33^{+0.05}_{-0.07}$ \\
SM~\cite{zgrad}   & 0.501 & 0.192 & -0.502  & -0.347 \\ \hline
\hline
\end{tabular}
\label{Tab:zquark}
\end{center}
\end{table*}

\begin{table*}
\begin{center}
\caption{The correlation coefficients between $g_V^{u}$, $g_A^u$, $g_V^d$ and
$g_A^d$. Only half of the symmetric correlation matrix is presented.}
\begin{tabular}{l|cr@{$.$}lr@{$.$}lc}
\hline
\hline
         & $g_A^{u}$ & \multicolumn{2}{c}{$g_V^{u}$} & \multicolumn{2}{c}{$g_A^{d}$} & $g_V^{d}$ \\ \hline
$g_A^{u}$ & 1.000   & \multicolumn{2}{c}{}          &  \multicolumn{2}{c}{}         & \\
$g_V^{u}$ & 0.470   & 1&000                         &  \multicolumn{2}{c}{}         & \\
$g_A^{d}$ & 0.201   & -0&606                        & 1&000                         & \\
$g_V^{d}$ & 0.217   & 0&925                         & -0&813                        & 1.000 \\
\hline
\end{tabular}
\label{Tab:coefficient}
\end{center}
\end{table*}

\section{Conclusions}
\indent We have measured the forward-backward charge
asymmetry in $p\bar{p}\rightarrow Z/\gamma^* \rightarrow e^+e^-$
events and extracted $\stwefflep$, $g_V^{u(d)}$ and $g_A^{u(d)}$ using
5.0~fb$^{-1}$ of integrated luminosity collected by the D0 experiment at $\sqrt{s}=1.96$~TeV. The measured
forward-backward charge asymmetry in the range $50<M_{ee}<1000$~GeV
agrees with the theoretical predictions. The measured $\stwefflep$ value 
can be directly compared with the LEP and SLD results, and
the overall $\stwefflep$ uncertainty for light quarks obtained is smaller than the
combined uncertainty in the LEP measurements of the inclusive
hadronic charge asymmetry. We also present the most precise direct
measurement to date of $g_V^{u}$, $g_A^{u}$, $g_V^{d}$, and $g_A^{d}$.

Although the uncertainty of our $\stwefflep$
measurement is still larger than that of the current world average, 
with about 10 fb$^{-1}$ of integrated luminosity expected by the end of Tevatron Run II, 
a combined result of CDF and D0 $A_{FB}$ measurements 
in both dielectron and dimuon channels has the potential to 
substantially impact the world average value of $\stwefflep$. 
In addition to a reduction of the dominant statistical uncertainties, 
many of the systematic uncertainties 
have a strong statistical component or will decrease with higher statistics,
for example the electron energy scale and resolution. 
To match the precision of the current world average of $\stwefflep$, 
the theoretical uncertainty due to PDFs need to be reduced in similar proportions 
as the experimental uncertainties of the measurement.

\section{Acknowledgements}
%
We thank the staffs at Fermilab and collaborating institutions,
and acknowledge support from the
DOE and NSF (USA);
CEA and CNRS/IN2P3 (France);
FASI, Rosatom and RFBR (Russia);
CNPq, FAPERJ, FAPESP and FUNDUNESP (Brazil);
DAE and DST (India);
Colciencias (Colombia);
CONACyT (Mexico);
KRF and KOSEF (Korea);
CONICET and UBACyT (Argentina);
FOM (The Netherlands);
STFC and the Royal Society (United Kingdom);
MSMT and GACR (Czech Republic);
CRC Program and NSERC (Canada);
BMBF and DFG (Germany);
SFI (Ireland);
The Swedish Research Council (Sweden);
and
CAS and CNSF (China).
%


\begin{thebibliography}{99}
\vskip 0.25cm

\bibitem{pdg}
 C. Amsler {\it et al.} (Particle Data Group), Phys. Lett. B {\bf 667}, 1 (2008) and 2009 partial update for the 2010 edition

\bibitem{cs_frame}
 J. C. Collins and D. E. Soper, Phys. Rev. D {\bf 16}, 2219 (1977).

\bibitem{cs_frame_plot}
 R. Gelhaus, FERMILAB-THESIS-2005-22, UMI-31-91669 (2005).

\bibitem{pythia}
 T. Sj$\ddot{\text o}$strand {\it et al.}, Comput. Phys. Commun. {\bf 135}, 238 (2001). {\sc pythia} version v6.323 is used throughout.

\bibitem{cteq}
  J. Pumplin {\it et al.}, J. High Energy Phys. {\bf 07}, 012 (2002); D. Stump {\it et al.}, J. High Energy Phys. {\bf 10}, 046 (2003).

\bibitem{zprime}
  J. L. Rosner, Phys. Rev. D {\bf 54}, 1078 (1996); M. Carena, A. Daleo, B. A. Dobrescu, T. M. P. Tait, Phys. Rev. D {\bf 70}, 093009 (2004).

\bibitem{led}
  H. Davoudiasl, J. L. Hewett, and T. G. Rizzo, Phys. Rev. Lett. {\bf 84}, 2080 (2000).

\bibitem{highmass}
  A. Abulencia {\it et al.} (CDF Collaboration), Phys. Rev. Lett. {\bf 95}, 252001 (2005).

\bibitem{highmass_1}
  D. Acosta {\it et al.} (CDF Collaboration), Phys. Rev. Lett. {\bf 95}, 131801 (2005).

\bibitem{highmass_2}
  V. M. Abazov {\it et al.} (D0 Collaboration), Phys. Rev. Lett. {\bf 95}, 091801 (2005).

\bibitem{highmass_3}
  V. M. Abazov {\it et al.} (D0 Collaboration), Phys. Rev. Lett. {\bf 95}, 161602 (2005).

\bibitem{highmass_4}
  T. Aaltonen {\it et al.} (CDF Collaboration), Phys. Rev. Lett. {\bf 99}, 171802 (2007).

\bibitem{highmass_5}
  V. M. Abazov {\it et al.} (D0 Collaboration), Phys. Rev Lett. {\bf 100}, 091802 (2008).

\bibitem{highmass_CDF}
  A. Abulencia {\it et al.} (CDF Collaboration), Phys. Rev. Lett. {\bf 96}, 211801 (2006).


\bibitem{lep_sinthetaW}
  G. Abbiendi {\it et al.} (LEP Collaborations ALEPH, DELPHI, L3 and OPAL; SLD Collaboration,
LEP Electroweak Working Group, SLD Electroweak and Heavy Flavor
Groups), Phys. Rep. {\bf 427}, 257 (2006).

\bibitem{zgrad}
  U. Baur, S. Keller, and W. K. Sakumoto, Phys. Rev. D {\bf 57}, 199 (1998); U. Baur, O. Brein, W. Hollik, C. Schappacher, D. Wackeroth, Phys. Rev. D {\bf 65}, 033007 (2002).

\bibitem{resbos}
  C. Balazs and C. P. Yuan, Phys. Rev. D {\bf 56}, 5558 (1997).

\bibitem{atomic_PV}
  S. C. Bennett and C. E. Wieman, Phys. Rev. Lett. {\bf 82}, 2484 (1999).

\bibitem{Moller}
 P. L. Anthony {\it et al.} (SLAC E158 Collaboration), Phys. Rev. Lett. {\bf 95}, 081601 (2005).

\bibitem{NuTeV}
  G. P. Zeller {\it et al.} (NuTeV Collaboration), Phys. Rev. Lett. {\bf 88}, 091802 (2002) [Erratum-ibid. {\bf 90}, 239902 (2003)].

\bibitem{cdf_RunII}
 D. Acosta {\it et al.} (CDF Collaboration), Phys. Rev. D {\bf 71}, 052002 (2005).

\bibitem{H1}
 A. Aktas {\it et al.} (H1 Collaboration), Phys. Lett. B {\bf 632}, 35 (2006).

\bibitem{cdf_RunI}
 T. Affolder {\it et al.} (CDF Collaboration), Phys. Rev. Lett. {\bf 87}, 131802 (2001); F. Abe {\it et al.} (CDF Collaboration), Phys. Rev. Lett. {\bf 77}, 2616 (1996).

\bibitem{d0_RunI}
 B. Abbott {\it et al.} (D0 Collaboration), Phys. Rev. Lett. {\bf 82}, 4769 (1999).

\bibitem{d0_RunII}
 V. M. Abazov {\it et al.} (D0 Collaboration), Phys. Rev. Lett. {\bf 101}, 191801 (2008).

\bibitem{d0lumi}
 T.~Andeen {\it et al.}, FERMILAB-TM-2365 (2007).

\bibitem{d0det}
 V. M. Abazov {\it et al.} (D0 Collaboration), Nucl. Instrum. Methods Phys. Res. Sect. A {\bf 565}, 463 (2006).

\bibitem{d0_coordinate}
D0 uses a cylindrical coordinate system with the $z$ axis running
along the beam axis in the proton direction. Angles $\theta$ and $\phi$ are the polar and
azimuthal angles, respectively. Pseudorapidity is defined as
$\eta=-\ln [\tan(\theta/2)]$ where $\theta$ is measured with respect
to the interaction vertex. In the massless limit, $\eta$ is equivalent to
the rapidity $y=(1/2) \ln[(E+p_z)/(E-p_z)]$
and $\eta_{\text{det}}$ is the pseudorapidity measured with respect to the
center of the detector. 

\bibitem{geant}
  R. Brun and F. Carminati, CERN Program Library Long Writeup
  W5013, 1993 (unpublished).

\bibitem{kfactor}
  P. B. Arnold, M. H. Reno, Nucl. Phys. {\bf B 319}, 37 (1989) [Erratum-ibid. {\bf B 330}, 284 (1990)].

\bibitem{NLO_corr}
  R. Hamberg, W. L. van Neerven, and T. Matsuura, Nucl. Phys. {\bf B 359}, 343 (1991); [Erratum-ibid. {\bf B 644}, 403 (2002)].

\bibitem{WW_NLO_corr}
  J. M. Campbell and R. K. Ellis, Phys. Rev. D {\bf 60}, 113006 (1999).

\bibitem{ttbar_NLO_corr}
  N. Kidonakis and R. Vogt, Phys. Rev. D {\bf 68}, 114014 (2003); M. Cacciari {\it et al.}, J. High Energy Phys. {\bf 04}, 68 (2004).

\bibitem{matrix_inversion}
  G. L. Marchuk, Methods of Numerical Mathematics (Springer, Berlin, 1975).

\bibitem{alpgen_d0}
  M. L. Mangano {\it et al.}, J. High Energy Phys. {\bf 07}, 001 (2003).

\bibitem{zfitter}
 D. Y. Bardin {\it et al.}, Z. Phys. C {\bf 44} 493 (1989);
 D. Y. Bardin {\it et al.}, Comput. Phys. Commun. {\bf 59}, 303 (1990).

\bibitem{H1_2010}
Z. Zhang, PoS DIS2010, 056 (2010).

\bibitem{minuitfit}
F. James, CERN Program Program Library Long Writeup D506, 1993 (unpublished)

\end{thebibliography}
\end{document}